\documentclass[conference,letterpaper,10pt]{IEEEtran}   

\usepackage{preamble}

\title{On the Rate-Distortion Function for Sampled Cyclostationary Gaussian Processes with Memory: Extended Version with Proofs\vspace{-0.2cm}}

\author{Zikun Tan, Ron Dabora, H. Vincent Poor \vspace{-2.8cm}
\thanks{Z. Tan and R. Dabora are with the School of ECE, Ben-Gurion University, Be'er-Sheva, Israel (emails: tanziku@post.bgu.ac.il, daborona@bgu.ac.il). R. Dabora is currently a Visiting Fellow at Princeton University. H. V. Poor is with the Department of ECE, Princeton University, Princeton, NJ, USA (email: poor@princeton.edu). This work was supported in part by the Israel Science Foundation under Grant 584/20, and in part by the U.S National Science Foundation under Grants CNS-2128448 and ECCS-2335876.}
}

\begin{document}   

\pagestyle{plain}

\maketitle   

\begin{abstract}   
In this work we study the \gls{rdf}  for lossy compression of
asynchronously-sampled \gls{ct} \gls{wscs} Gaussian processes with memory. As the case of
synchronous sampling, i.e., when the sampling interval is commensurate with the period of the cyclostationary statistics, has already been studied, we focus on   \gls{dt} processes obtained by asynchronous sampling, i.e., when the sampling interval is incommensurate with the period of the cyclostationary statistics of the \gls{ct} \gls{wscs} source process. It is further assumed that the sampling interval is smaller than the maximal autocorrelation length of the \gls{ct} source process, which implies that the \gls{dt} process possesses memory. Thus, the sampled process is a \gls{dt} \gls{wsacs}  processes with memory. This problem is motivated by the fact that man-made communications signals are modelled as \gls{ct} \gls{wscs} processes; hence, applications of such sampling include, e.g., compress-and-forward relaying and recording systems. The main challenge follows because, with asynchronous sampling, the \gls{dt} sampled process is not information-stable, and hence the characterization of its \gls{rdf} should be carried out within the information-spectrum framework instead of using conventional information-theoretic arguments. This work expands upon our previous work which addressed the special case in which the \gls{dt} process is independent across time. The existence of dependence between the samples requires new tools to obtain the characterization of the \gls{rdf}.
\end{abstract}

\glsresetall   

\section{Introduction}   
\label{introduction}
As man-made sequences and waveforms typically involve  repetitive operations, many communications signals inherently possess periodically time-varying statistical characteristics, and hence are commonly modeled as {\gls{ct} \gls{wscs} processes} \cite[Sec.~I]{gardner2006}. Modern communications systems process \gls{dt} signals, which are obtained by sampling the \gls{ct} received signals. Sampling a \gls{ct} \gls{wscs} signal does \emph{not} necessarily result in a \gls{dt} \gls{wscs} signal: if the ratio between the sampling interval and the period of the autocorrelation function of the \gls{ct} \gls{wscs} process is a \emph{rational number}, a situation  referred to as \emph{synchronous sampling}, the sampled process is a \gls{dt} \gls{wscs} process; however, if this ratio is an \emph{irrational number}, which is referred to as \emph{asynchronous sampling}, the sampled process is a \gls{dt} \gls{wsacs} process\cite[Sec.~3]{izzo1996}, \cite[Sec.~3.9]{gardner2006}. Asynchronous sampling could arise, for example, in relay channels, in which the relay employs  compress-and-forward  \cite{cover1979}: the relay first compresses its DT sampled received signal and subsequently forwards the compressed version to the destination \cite{dabora2008}. Practically, due to  clock jitter, see, e.g., \cite{vig1993, azeredo-leme2011}, the sampling interval at the relay and the symbol interval of the \gls{ct} received signal would be incommensurate, resulting in asynchronous sampling. Furthermore, to decrease the loss of information associated with sampling the \gls{ct} received signal at the relay, the sampling interval is typically selected to be shorter than the maximal autocorrelation length of the received \gls{ct} process, giving rise to a correlated \gls{dt} sampled process. We also note that in many practical scenarios, communications signals can be (asymptotically) modeled as Gaussian processes. Such scenarios include \gls{ofdm} signals with sufficiently long sequences of symbols 
\cite{wei2010}, and signals received over \gls{lti} channels with long memories \cite[Sec.~III-A]{metzger1987}. It thus follows that in many communications scenarios, the \gls{dt} sampled received signal can be modelled as a \gls{dt} \gls{wsacs} Gaussian process with memory.

In this work, we study the rate-distortion characterization for the lossy compression of \gls{dt} \gls{wsacs} Gaussian processes with memory. Nonstationarity and nonergodicity of such processes imply that these processes are \emph{not} information-stable, and thus conventional information-theoretic arguments relying on the \gls{aep} (see \cite[Ch.~3]{cover2006}) cannot be applied. Instead, we use the \emph{information-spectrum framework} \cite{han1993,han2010}, which supports concepts that facilitate the derivation of the \gls{rdf} for information-unstable processes. These concepts were successfully applied in the study  of  the dual problem, of capacity characterization for channels with additive \gls{dt}  \gls{wsacs} Gaussian noise, in \cite{shlezinger2020} for the memoryless case and in \cite{dabora2023} for the case with memory.

The \gls{rdf} for the lossy compression of \gls{dt} \gls{wscs} Gaussian processes with memory was derived in \cite{kipnis2018} by applying the \gls{dcd} \cite[Rep.~1]{giannakis1999}, which transforms a DT WSCS process into an equivalent multi-dimensional DT stationary process. In \cite{abakasanga2020}, the \gls{rdf} for compressing \gls{dt} memoryless \gls{wsacs} Gaussian  processes was derived, using the information-spectrum framework. To the best of our knowledge, the \gls{rdf} for compressing \gls{dt} \gls{wsacs} Gaussian processes with memory has not been characterized to date, which is the focus of this work.

\tib{Main Contributions:} In this work, we derive the \gls{rdf} for the lossy compression of \gls{dt} \gls{wsacs} Gaussian processes with memory, obtained by asynchronously sampling \gls{ct} \gls{wscs} Gaussian source processes using a sampling interval smaller than the maximal autocorrelation length of the \gls{ct} source process, subject to the \gls{mse} distortion. Due to the information-instability of the \gls{dt} \gls{wsacs} process, the \gls{rdf} characterization is carried out within the information-spectrum framework. We derive the \gls{rdf} subject to the assumptions that the initial sampling phase is synchronized at the encoder and at the decoder, while the source emits messages constantly and without delay between subsequent source messages.

\tib{Organization:} The rest of this work is organized as follows: Sec.~\ref{preliminary background} presents a theoretical review  of \gls{wscs} processes,
of synchronous sampling and of asynchronous sampling applied to
\gls{ct} \gls{wscs} processes, and of rate-distortion theory; Sec.~\ref{problem formulation and model} formulates the problem and states the model of the source process for sampling and compression; Sec.~\ref{results} presents the main result of this work as well as a discussion of its implications; Sec.~\ref{conclusion} concludes the  work.

\section{Preliminary Background}   
\label{preliminary background}

\subsection{Notations}
\label{notations and symbols}
Throughout this work, we denote the sets of real numbers, positive real numbers, rational numbers, integers, non-negative integers and positive integers by $\mR$, $\mRdplus$, $\mQ$,  $\mZ$, $\mN$, and $\mNplus$, respectively. Scalar \glspl{rv} (resp. deterministic values) are denoted with uppercase letters, e.g., $X$ (resp., lowercase letters, e.g., $x$). Round brackets denote \gls{ct} random processes and deterministic functions, e.g., $X(t)$, $t\in \mR$, denotes a \gls{ct} random process, and $x(t)$,  $t\in \mR$, denotes a \gls{ct} deterministic function. Square brackets denote \gls{dt} random processes and deterministic functions, e.g., $X[i]$, $i\in \mZ$, denotes a \gls{dt} random process and $x[i]$, $i\in \mZ$, denotes a \gls{dt} deterministic function. Matrices are denoted with sans serif uppercase letters, e.g., $\mmat{A}$, and the element at the $i$-th row and $j$-th column of the matrix $\mmat{A}$ is denoted with $(\mmat{A})_{i,j}$.
Transpose, inverse, rank, range and null-space of a matrix $\mmat{A}$ are denoted by $\mmat{A}^T$, $\mmat{A}^{-1}$, $\rank(\mmat{A})$, $\range(\mmat{A})$ and $\nulspc(\mmat{A})$, respectively. For a $k\times k$ square matrix $\mmat{B}$, $\det(\mmat{B})$ denotes its determinant, $\tr\{\mmat{B}\}$ denotes its trace and $\mmat{B}\succcurlyeq 0$ (resp., $\mmat{B}\succ 0$) denotes it is positive semidefinite (resp., positive definite). We denote a $k\times k$ identity matrix, a $k\times l$ zero matrix and a $k\times k$ square zero matrix as $\idmat_{k}$, $\zeromat_{k\times l}$ and $\zeromat_{k}$, respectively. Boldface uppercase (resp., lowercase) letters denote column random (resp., deterministic) vectors, e.g., $\mvec{X}$ (resp., $\mvec{x}$). When the length of the vector needs to be explicitly stated, it is denoted as $X^{(l)}\equiv\mvec{X}$ (resp., $x^{(l)}\equiv\mvec{x}$) for random (resp., deterministic) vectors of length $l$. $\zerovec_{k}$ represents an all-zero column vector of length $k$. The autocorrelation matrix of a column random vector $\mvec{X}$ is denoted as $\mmat{C}_{\mvec{X}}$. $\mvec{X}\sim\mGd(\mvlgrk{\mu}_{\mvec{X}}, \mmat{K}_{\mvec{X}})$ denotes a real Gaussian random vector with a mean vector $\mvlgrk{\mu}_{\mvec{X}}$ and an autocovariance matrix $\mmat{K}_{\mvec{X}}$. $\mE\{\cdot\}$ denotes the expectation, $\var\{\cdot\}$ denotes the variance, $\eqdist$ represents equality in distribution, $|\cdot|$ denotes the absolute value, $\lceil \cdot \rceil$ represents the ceiling function, $\log(\cdot)$ denotes the base-$2$ logarithm function, and $\pr\{\cdot\}$ denotes the probability distribution. The differential entropy of a real \gls{rv} $X$ and the mutual information of a pair of real \glspl{rv} $X$ and $Y$ are denoted with $h(X)$ and $I(X;Y)$, respectively.

\subsection{Wide-Sense Cyclostationary Processes}
\label{wide-sense cyclostationary processes}
We next briefly review some definitions regarding cyclostationary processes. We begin by recalling the definition of \gls{wscs} processes as follows:

\begin{definition}[\gls{wscs} processes {\cite[Def.~17.1]{giannakis1999}, \cite[Sec.~3.2]{gardner2006}}]
A real-valued \gls{ct} (resp., \gls{dt}) random process $X(t)$, $t \in \mR$ (resp., $X[i]$, $i \in \mZ$) is called \emph{\gls{wscs}} if both its mean $m_X(t)$ (resp., $m_X[i]$) and its autocorrelation function $c_X(t,\lambda)$ (resp., $c_X[i,\Delta]$) are periodic in time $t$ (resp., time index $i$) with some period $T_c \in \mRdplus$ (resp., $N_c \in \mN^+$) for any lag $\lambda \in \mR$ (resp., $\Delta \in \mZ$).
\end{definition}

The definition of \gls{wsacs} processes requires first to recall the definition of \gls{dt} almost periodic functions,  stated below:

\begin{definition}[\gls{dt} almost periodic functions\footnote{There are several different definitions of almost periodic functions. This defined function is almost periodic in the sense of Bohr (see \cite{bohr2018}).} {\cite[Def.~2.1]{cherif2011}, \cite[Def.~11]{guan2013}}]
\label{definition of almost periodic functions}
A real-valued \gls{dt} deterministic function $f[i]$, $i\in\mZ$, is called \emph{almost periodic}, if for any $\epsilon \in \mRdplus$, there exists an associated number  $\tilde{l_{\epsilon}}\in\mNplus$ which satisfies that for any $\beta\in\mZ$, there exists $\Delta\in[\beta, \beta+\tilde{l_{\epsilon}}]$, such that $\sup_{i\in\mZ} |f[i+\Delta]-f[i]|<\epsilon$.
\end{definition}

Lastly, using Def.~\ref{definition of almost periodic functions} we present the definition of \gls{dt} \gls{wsacs} processes as follows:

\begin{definition}[\gls{dt} \gls{wsacs} processes {\cite[Def.~17.2]{giannakis1999}, \cite[Sec.~3.2]{gardner2006}}]
A real-valued \gls{dt} random process $X[i]$, $i \in \mZ$, is called \emph{\gls{wsacs}} if both its mean $m_X[i]$ and its autocorrelation function  $c_X[i,\Delta]$ are \emph{almost periodic} in time index $i$ for any lag value $\Delta \in \mZ$.
\end{definition}

It is noted that communications signals are often zero-mean in practice \cite[Sec.~17.2]{giannakis1999}, and thus the
cyclostationarity periods of these signals are determined by the periods of their autocorrelation functions.

\subsection{Discrete-Time Processes Obtained by Sampling Continuous-Time Wide-Sense Cyclostationary Processes}
\label{subsec: sampling effects}
In this subsection, we briefly recall the discussion in \cite[Sec.~II-B]{shlezinger2020} on the statistics of sampled \gls{ct} \gls{wscs} processes.

Consider sampling a \gls{ct} \gls{wscs} process $X_c(t)$ whose cyclostationarity period is $T_c$, with a sampling interval $T_s$ and a sampling phase $\phi_s \in [0, T_c)$. Let $X[i|T_s,\phi_s] \triangleq X_c(i \cdot T_s+\phi_s)$ denote the \gls{dt} sampled process. As noted in Sec.~\ref{introduction}, the \gls{dt} process obtained by sampling a \gls{ct} \gls{wscs} process may not be a \gls{dt} \gls{wscs} process. In fact, the statistical characteristics of such \gls{dt} sampled processes are strongly dependent on the value of $\phi_s$ and the relationship between $T_s$ and $T_c$. Specifically, when $T_s/T_c$ is a rational number then the sampling is said to be {\em synchronous}, and the resulting \gls{dt} process is \gls{wscs}. However, when $T_s/T_c$ is an irrational number then the sampling is said to be {\em asynchronous}, and the resulting \gls{dt} process is \gls{wsacs}. Such dependence is clearly different from the sampling of \gls{ct} {\em stationary} processes, which {\em always} results in \gls{dt} stationary processes \cite[Sec.~II-B]{shlezinger2020}. Consequently, the common practice of applying signal processing algorithms designed for stationary signals in processing cyclostationary signals often leads to erroneous results, e.g., incorrect formulas for \glspl{psd}, see \cite{gardner1987}.

To numerically demonstrate this point (see also \cite[Sec.~II-B]{shlezinger2020}), let us assume that the process $X_c(t)$ has a variance function given by $\var\big(X_c(t)\big) =\frac{1}{3} \cdot \sin(\frac{2\pi t}{T_c})+5$ and consider three sampling scenarios: For the first sampling scenario, the process $X_c(t)$ is sampled uniformly with $T_s = \frac{T_c}{4}$ and $\phi_s=0$. The resulting  \gls{dt} sampled process has a  variance function given by $\var\big(X[i]\big|\frac{T_c}{4},0\big) = \var\big(X_c(i \cdot \frac{T_s}{4})\big) = \{5, 5.333, 5, 4.667, 5, 5.333, 5, 4.667, \cdots \}$~for $i = 0, 1, 2, 3, 4, 5, 6, 7, \cdots$. It can be observed that this sampled variance function is periodic with  period of $4$, hence the corresponding  \gls{dt} sampled process is \gls{wscs}. For the second sampling scenario let the process $X_c(t)$ be sampled uniformly with $T_s = \frac{T_c}{4}$ and $\phi_s=\frac{\pi}{3}$. The resulting \gls{dt} variance is $\var\big(X[i]\big|\frac{T_c}{4},\frac{\pi}{3}\big) =\var\big(X_c(i \cdot \frac{T_s}{4}+\frac{\pi}{3})\big) = \{5.332, 4.975, 4.668, 5.025, 5.332, 4.975, 4.668, 5.025, \cdots \}$~for $i = 0, 1, 2, 3, 4, 5, 6, 7, \cdots$. It is observed that this \gls{dt} variance function is also periodic with  period of $4$, but with {\em different values} within each period compared to the first sampling case, due to the different sampling phase. Hence, both \gls{dt} sampled processes obtained for the two scenarios are \gls{dt} \gls{wscs} processes with a cyclostationarity period of $4$, yet the variance functions of these two \gls{dt} processes are different, namely, these are two \emph{different} \gls{dt} \gls{wscs} processes. We note that in both sampling scenarios above, $T_s/T_c$ is a {rational} number, which corresponds to {synchronous sampling}. For the last scenario, consider $T_s=\frac{T_c}{4\pi}$ and $\phi_s=\frac{\pi}{3}$. The variance function of the \gls{dt} sampled process is $\var\big(X[i]\big|\frac{T_c}{4\pi},\frac{\pi}{3}\big) = \var\big(X_c(i \cdot \frac{T_c}{4\pi}+\frac{\pi}{3})\big) = \{ 5.167, 5.285, 5.333, 5.300, 5.193, 5.039,$ $4.876, 4.743, \cdots \}$,~for $i = 0, 1, 2, 3, 4, 5, 6, 7, \cdots$. This \gls{dt} variance function is not periodic but is \emph{almost periodic}, hence, the corresponding \gls{dt} sampled process is \gls{wsacs}. We note that in this case the ratio  $T_s/T_c$ is {irrational}, which corresponds to {asynchronous sampling}. These sampling cases demonstrate that both the sampling interval and the sampling phase strongly impact the statistical characteristics of  \gls{dt} processes obtained by sampling \gls{ct} \gls{wscs} processes and therefore necessarily affect the \gls{rdf} characterization of such \gls{dt} processes.

\subsection{The Rate-Distortion Function}   
\label{rate-distortion theory}
In the following, we review several relevant definitions and notions related to  \glspl{rdf}, beginning with the definition of a lossy source code:

\begin{definition}[Lossy source codes {\cite[Sec.~10.2]{cover2006},\cite[Sec.~3.6]{el_gamal2011}}]
A \emph{lossy source code} $(M,l)$ with a message set of size $M$ and a \emph{blocklength} of  $l$ consists of:
\begin{itemize}
    \item an \emph{encoder} $f_l(\cdot)$ that maps a sequence of $l$ source symbols $\{x_i\}_{i=0}^{l-1} \equiv x^l$, over corresponding alphabets $\{\malp{X}_i\}_{i=0}^{l-1} \equiv \malp{X}^l$, into an index selected from a message set containing $M$ indices, $f_l(\cdot): \malp{X}^l \mapsto \{1,2,\ldots,M\}$,
    \item a \emph{decoder} $g_l(\cdot)$ that assigns a sequence of $l$ reconstruction symbols $\{\mr{x}_i\}_{i=0}^{l-1} \equiv \mr{x}^l$, over corresponding alphabets $\{\mralp{X}_i\}_{i=0}^{l-1} \equiv \mralp{X}^l$, to each received message index, $g_l(\cdot): \{1,2,\ldots,M\} \mapsto \mralp{X}^l$,
\end{itemize}
\end{definition}
where the set $\big\{g_l(i)\big\}_{i=1}^{M}$, which corresponds to $\big\{\mr{X}^{l}(i)\big\}_{i=1}^{M}$, constitutes the \emph{codebook} and $\{f_n^{-1}(i)\}_{i=1}^{M}$ are the associated \emph{assignment regions} \cite[Sec.~10.2]{cover2006}.
The \emph{code rate}  of such a code is defined as $R\triangleq\frac{1}{l}\log M$.

In lossy source coding, the error between a sequence of source symbols and the corresponding sequence of reconstruction symbols is characterized using a \emph{distortion function} $d(\cdot,\cdot)$, which is defined on a symbol-by-symbol basis and maps each pair of the source symbol alphabet $\malp{X}$ and \tred{its} corresponding reconstruction symbol alphabet $\mralp{X}$ into the set of non-negative real numbers \cite[Sec.~10.2]{cover2006}, \cite[Sec.~3.6]{el_gamal2011}, i.e., $d(\cdot,\cdot): \malp{X} \times \mralp{X} \mapsto \mR^+$.
Then, the distortion function for a sequence of source symbols and its corresponding sequence of reconstruction symbols is the arithmetic mean of the per-symbol distortion values:
\begin{equation*}
    \maa{d}(x^l,\mr{x}^l) \triangleq \frac{1}{l} \sum_{i=0}^{l-1} d\big(x[i],\mr{x}[i]\big).
\end{equation*}

\vspace{-0.3cm}
A commonly used distortion function for compressing continuous sources is the squared-error distortion function, which is defined as
\begin{equation*}
    d_{se}(x,\mr{x}) \triangleq (x-\mr{x})^2.
\end{equation*}

We say a rate-distortion pair $(R,D)$ is \emph{achievable}, if it satisfies the following definition:

\begin{definition}[Achievable rate-distortion pairs {\cite[Sec.~10.2]{cover2006}, \cite[Sec.~3.6]{el_gamal2011}}]
The rate-distortion pair $(R,D)$ is said to be \emph{achievable} if there exists a sequence of lossy source codes $(2^{lR},l)$ for which
\begin{equation*}
    \limsup_{l\to\infty} \mE\Big\{\maa{d}\Big(X^l,g_l\big(f_l(X^l)\big)\Big)\Big\} \leq D.
\end{equation*}
\end{definition}

Finally, the \gls{rdf}, denoted as $R(D)$, is defined as follows:

\begin{definition}[\gls{rdf} {\cite[Sec.~IV-A]{berger1998}, \cite[Sec.~10.2]{cover2006}, \cite[Sec.~3.6]{el_gamal2011}}]
The \gls{rdf} $R(D)$ is the infimum of all code rates $R$ for which the rate-distortion pair $(R,D)$ is achievable, for a given average distortion constraint $D$.
\end{definition}

\section{Problem Formulation and Model}   
\label{problem formulation and model}
We consider a real-valued \gls{ct} \gls{wscs} Gaussian source process $X_c(t)$, with an autocorrelation function $c_{X_c}(t, \lambda) \triangleq \mE\{X_c(t) \cdot X_c(t+\lambda)\}$, where $\lambda$ is the lag. The function $c_{X_c}(t, \lambda)$ is assumed to be \emph{bounded} and \emph{uniformly continuous} in both $t$ and $\lambda$, and is periodic in $t$ with  period  $T_c \in \mRdplus$, i.e., $c_{X_c}(t, \lambda)=c_{X_c}(t+T_c, \lambda)$, $\forall t, \lambda \in \mR$. It is also assumed that $X_c(t)$ is a \emph{finite-memory} process, in the sense that it has a finite maximal autocorrelation length $\lambda_c \in \mRdplus$, i.e., $c_{X_c}(t, \lambda)=0$, $\forall |\lambda| > \lambda_c$.

The \gls{ct} source process $X_c(t)$ is sampled uniformly with a sampling interval $T_s(\epsilon)=\frac{T_c}{p+\epsilon}$, where $p \in \mNplus$ and $\epsilon \in [0,1)$ represents the \emph{synchronization mismatch} in the sampling interval. The resulting \gls{dt} sampled source process is denoted by $X_{\epsilon}^{\phi_s}[i]=X_c\big(i \cdot T_s(\epsilon)+\phi_s\big)$, where $\phi_s \in [0,T_c)$ denotes the sampling phase. Since in this work we assume that the sampling interval $T_s(\epsilon)$ is smaller than the maximal autocorrelation length $\lambda_c$ of the \gls{ct} source process, the resulting \gls{dt} sampled source process $X_{\epsilon}^{\phi_s}[i]$ has memory. The autocorrelation function for $X_{\epsilon}^{\phi_s}[i]$, $c_{X_{\epsilon}^{\phi_s}}[i,\Delta]$, at time index $i$ and lag $\Delta$, is given as (see also \cite[Sec. III-A]{dabora2023})
\vspace{-0.1cm}
\begin{align}
\label{eqn: af of sampled source process}
    & c_{X_{\epsilon}^{\phi_s}}[i,\Delta] \nonumber \\
    & \triangleq \mE\{X_{\epsilon}^{\phi_s}[i] \cdot X_{\epsilon}^{\phi_s}[i+\Delta]\} \nonumber\\
    & = \mE\big\{X_c\big(i \cdot T_s(\epsilon)+\phi_s\big) \cdot X_c\big((i+\Delta) \cdot T_s(\epsilon)+\phi_s\big)\big\} \nonumber\\
    & = c_{X_c}\Big(\frac{i \cdot T_c}{p+\epsilon}+\phi_s, \frac{\Delta \cdot T_c}{p+\epsilon}\Big).
\end{align}

\vspace{-0.2cm}
Due to the maximal autocorrelation length $\lambda_c$ of the \gls{ct} source process $X_c(t)$, we have $c_{X_{\epsilon}}^{\phi_s}[i,\Delta]=0$, $\forall |\Delta| \geq \Big\lceil \frac{(p+1) \cdot \lambda_c}{T_c}\Big\rceil \triangleq \tau_c < \infty$, which means that $X_{\epsilon}^{\phi_s}[i]$ is also a \emph{finite-memory} process.

As discussed above, the ratio between $T_s(\epsilon)$ and $T_c$ determines the statistics of the \gls{dt} process obtained by sampling a \gls{ct} \gls{wscs} process: 
when $\epsilon \in \mQ$, i.e., $\exists u, v \in \mNplus$, s.t. $\epsilon = \frac{u}{v}$, the sampled process is a \gls{dt} \gls{wscs} process with a cyclostationarity period of $N_c = p \cdot v+u \triangleq p_{u,v}$ \cite[Sec.~II-C]{shlezinger2020}, which corresponds to synchronous sampling. However, when $\epsilon \notin \mQ$, namely $\epsilon$ cannot be expressed as a ratio of two positive integers, the resulting \gls{dt} sampled process is a \gls{dt} \gls{wsacs} process, which corresponds to asynchronous sampling.

In our scenario, the distortion between the \gls{dt} sampled source process and its reconstructed sequence at the receiver is measured via the \gls{mse} distortion function. Set the blocklength to $l$ and consider a sequence of source symbols $\{X_{\epsilon}^{\phi_s}[i]\}_{i-0}^{l-1}$ with its corresponding sequence of reconstruction symbols $\{\mr{X}_{\epsilon}^{\phi_s}[i]\}_{i-0}^{l-1}$. Let $\mvec{S}_{\epsilon,l}^{\phi_s} \triangleq \mvec{X}_{\epsilon,l}^{\phi_s} - \mr{\mvec{X}}_{\epsilon,l}^{\phi_s}$, where $\mvec{S}_{\epsilon,l}^{\phi_s} \equiv \{S_{\epsilon}^{\phi_s}[i]\}_{i-0}^{l-1}$. Using these definitions, the distortion constraint can be expressed as
\vspace{-0.1cm}
\begin{align}
    \label{a 1058am}
    & \nonumber \mE\Big\{\maa{d}_{se}\big(\{X_{\epsilon}^{\phi_s}[i]\}_{i-0}^{l-1},\{\mr{X}_{\epsilon}^{\phi_s}[i]\}_{i-0}^{l-1}\big)\Big\} \\
    & = \frac{1}{l} \cdot \mE\bigg\{\sum_{i=0}^{l-1} \big(X_{\epsilon}^{\phi_s}[i] - \mr{X}_{\epsilon}^{\phi_s}[i]\big)^{2}\bigg\} \notag\\
    & = \frac{1}{l} \cdot \tr\Big\{\mE\big\{\big(\mvec{X}_{\epsilon,l}^{\phi_s} - \mr{\mvec{X}}_{\epsilon,l}^{\phi_s}\big) \cdot \big(\mvec{X}_{\epsilon,l}^{\phi_s} - \mr{\mvec{X}}_{\epsilon,l}^{\phi_s}\big)^{T}\big\}\Big\} \notag\\
    & = \frac{1}{l} \cdot \tr\Big\{\mE\big\{\mvec{S}_{\epsilon,l}^{\phi_s} \cdot (\mvec{S}_{\epsilon,l}^{\phi_s})^{T}\big\}\Big\} \notag\\
    & = \frac{1}{l} \cdot \tr\big\{\mmat{C}_{\mvec{S}_{\epsilon,l}^{\phi_s}}\big\}.
\end{align}

\vspace{-0.2cm}
Next, we recall the definition of the limit superior in probability for a sequence of real \glspl{rv}:
\begin{definition}[Limit superior in probability {\cite[Def.~1.3.1]{han2010}}]
For a sequence of real \glspl{rv} $X_{i}$, $i\in\mN$, its \emph{limit superior in probability} is given as:
\vspace{-0.1cm}
\begin{equation*}
    \limsupp_{i\to\infty} {X_{i}} \triangleq \inf\big\{\alpha\in\mR | \lim_{i\to\infty} \Pr\{ X_{i} > \alpha \} = 0\big\} \triangleq \alpha_{0}.
\end{equation*}
\end{definition}
From this definition it follows that  $\forall \alpha<\alpha_{0}$ $\exists\mu\in\mRdplus$, s.t. there are countably many indices $i\in\mN$ for which $\Pr\{X_{i}>\alpha\} >\mu$.

Lastly, we present an auxiliary result on the distribution of quadratic Gaussian forms:
\begin{lemma}
    \label{Lemma:intergrability}
    Consider a sequence of $l\times 1$ random vectors $\{\Xvecl\}_{l\in\mN^+}$ where $\Xvecl\sim\mGd\big(\zerovec_l,\Cxl\big)$, 
    s.t. $\big(\Cxl\big)_{k,k}\le\beta<\infty$, $\forall 0\le k \le l-1$, $\forall l\in\mN^+$. 
    Then $W_l\triangleq \frac{1}{l}\big(\Xvecl\big)^T\cdot \Xvecl$ is uniformly integrable.
\end{lemma}
\begin{IEEEproof}
The proof is detailed in Appendix \ref{appendix:lemma_intefrability}.
\end{IEEEproof}

\section{Main Theorem and Discussion}
\label{results}
As detailed in Sec.~\ref{subsec: sampling effects}, the statistical properties of sampled \gls{ct} \gls{wscs} processes are determined by both the sampling interval and the sampling phase. Therefore, both  factors should be accounted for in the rate-distortion characterization. We assume that $T_c$ and $T_s(\epsilon)$ are known at both the source and the destination. In our analysis we also assume sampling phase synchronization between the encoder and the decoder w.r.t the period of the correlation, in the sense that when compressing each source message, both the encoder and the decoder know the sampling phase within $T_c$. This facilitates adapting the statistics of the source code according to the sampling phase to minimize the achievable code rate in real time. Note that such synchronization needs to be carried out only once, as $T_c$ and $T_s(\epsilon)$ are known at both the source and the destination. This knowledge enables the destination to independently keep track of the sampling phase after initial synchronization.
Assuming the source generates messages continuously and without delay between subsequent messages, the rate-distortion function for this scenario is characterized as follows:

\begin{theorem}
\label{thm: main theorem}
Consider a \gls{dt}  \gls{wsacs} Gaussian process  with memory, $X_{\epsilon}^{\phi_s}[i]$, obtained by uniformly sampling a \gls{ct} zero-mean \gls{wscs} Gaussian source process $X_c(t)$ whose correlation function  $c_{X_c}(t, \lambda)$ is periodic with period  $T_c\in\mRdplus$, has maximal autocorrelation length of $\lambda_c\in\mRdplus$, and a bounded variance, $c_{X_c}(t, 0)<\beta<\infty$, $\forall t\in\mR$. Let the sampling interval be $T_s(\epsilon)=\frac{T_c}{p+\epsilon} \leq \lambda_c$,  $\epsilon \in [0,1)$, $\epsilon\notin\mQ$, and the initial sampling phase $\phi_s \in [0,T_c)$. Let $\mC_{\mvec{S}_{\epsilon,l}^{\phi_s}} \triangleq \bigg\{\mmat{C}_{\mvec{S}_{\epsilon,l}^{\phi_s}}\in\mR^{l\times l}: \frac{1}{l}\tr\left\{\mmat{C}_{\mvec{S}_{\epsilon,l}^{\phi_s}}\right\} \leq D, \mmat{C}_{\mvec{X}_{\epsilon,l}^{\phi_s}} \succeq \mmat{C}_{\mvec{S}_{\epsilon,l}^{\phi_s}},
    \mmat{C}_{\mvec{S}_{\epsilon,l}^{\phi_s}} \succ 0,
    \mmat{C}_{\mvec{S}_{\epsilon,l}^{\phi_s}} = (\mmat{C}_{\mvec{S}_{\epsilon,l}^{\phi_s}})^{T} \bigg\}$.

If the initial sampling time w.r.t. $T_c$ is synchronized between the encoder and the decoder, then, for a given average distortion constraint $D$, the \gls{rdf} $R_{\epsilon}(D)$ is given as follows:
\begin{equation}
\label{a 0140am}
R_{\epsilon}(D) = \frac{1}{T_c} \int_{\phi_s=0}^{T_c} R_{\epsilon}^{\phi_s}(D) \mtxt{d} \phi_s
\vspace{-0.2cm}
\end{equation}
where
\vspace{-0.3cm}
\begin{multline}
    \label{eqn: rdf for given sampling phase}
    R_{\epsilon}^{\phi_s}(D) \triangleq 
    \limsup_{l\to\infty} \min_{ \mmat{C}_{\mvec{S}_{\epsilon,l}^{\phi_s}}\in \mC_{\mvec{S}_{\epsilon,l}^{\phi_s}}}  \frac{1}{2l} \log\bigg(\frac{\det(\mmat{C}_{\mvec{X}_{\epsilon,l}^{\phi_s}})}{\det(\mmat{C}_{\mvec{S}_{\epsilon,l}^{\phi_s}})}\bigg),
\end{multline}
which is achieved by $\mvec{S}_{\epsilon,l}^{\phi_s}\sim\mGd\big(\zerovec_l,\mmat{C}_{\mvec{S}_{\epsilon,l}^{\phi_s}}\big)$.
\end{theorem}

\begin{IEEEproof}
    The proof is detailed in Appendix \ref{Appendix:proof_of_main_theorem}.
\end{IEEEproof}

\begin{corollary}
    \label{corr:No_delay_fixed rate}
    If in the scenario in Thm.~\ref{thm: main theorem},  the code rate has to be fixed, for a given average distortion constraint $D$, the \gls{rdf} $R_{\epsilon}(D)$ is given as 
    \begin{equation}
    \label{eqn:RDF_fixed_rate_thm1}
        R_{\epsilon}(D) = \max_{\phi_s\in[0,T_c)} R_{\epsilon}^{\phi_s}(D),
    \end{equation}
    where $R_{\epsilon}^{\phi_s}(D)$ is given in Eqn.~\eqref{eqn: rdf for given sampling phase}.
\end{corollary}

\subsection{Discussion}   
\label{discussion}

\begin{remark}[A note on the use of a limit in Thm.~\ref{thm: main theorem} and in Corollary \ref{corr:No_delay_fixed rate}]
    For $\epsilon\notin\mQ$, it is possible to intuitively view the correlation function of the \gls{dt} process obtained by sampling as exhibiting all values of the \gls{ct} correlation function, given an asymptotically long sequence of samples. This follows because the \gls{dt} correlation function {\em never repeats} itself. For this reason, the \gls{rdf} characterization has to be stated as a limiting expression in the blocklength. 
    We note that while the expression contains a limit, still, our derivation obtains a {\em computable expression for each blocklength $l$}. 
   
    From the above intuition it follows that when sampling is asynchronous, the initial sampling phase $\phi_s$ should not affect the performance at asymptotically long blocks. In other words, we conjecture that
    \[
        R_{\epsilon}^{\phi_s}(D)  = R_{\epsilon}^{0}(D).
    \]
    Proving this conjecture  is a task we plan to include in our future work.

\end{remark}

 \begin{remark}[When sampling is synchronous]
    When $\epsilon\in\mQ$, the sampled \gls{dt} process is \gls{wscs}. This \gls{dt} process can be equivalently represented as a multivariate stationary process, thus, \gls{rdf} characterization can be obtained without taking the limit in $l$, see \cite{kipnis2018}. Then, if $\phi_s$ cannot be selected, the rate is determined by the initial phase at the beginning of transmission,
    \begin{equation*}
          R^{sync}_{\epsilon}(D) = R_{\epsilon}^{\phi_s}(D).
    \end{equation*}
        Alternatively, if $\phi_s$ can be selected for the initial transmission, then the \gls{rdf} can be characterized via the sampling phase $\phi_s$ which minimizes the rate, i.e. the rate is fixed and equal to
    \begin{equation*}
          R^{sync}_{\epsilon}(D) = \min_{\phi_s\in[0,T_c)} R_{\epsilon}^{\phi_s}(D).
    \end{equation*}    
    Note that in both \glspl{rdf} above, the limit in $l$ at $R_{\epsilon}^{\phi_s}(D)$ can be evaluated using a finite computation, see \cite{kipnis2018}.
\end{remark}

\begin{remark}[The relationship with the dual problem]
    The restriction $c_{X_c}(t, 0)<\beta<\infty$ is required to guarantee that the distortion, w.r.t. a reference reconstruction word $\hxvecl=\zerovec_l$ is uniformly integrable, see \cite[Eqn. (5.3.4)]{han2010}. This, in turn, is required by the achievability argument in the proof of Thm.~\ref{thm: main theorem}, which applies \cite[Thm. 5.5.1]{han2010} to conclude that the limit-superior in probability is indeed a lower bound on the code rate subject to an average distortion constraint. This restriction highlights a fundamental difference between the current work and the study of the dual problem, of the capacity of \gls{dt} communications channels with additive \gls{wsacs} Gaussian noise with memory, considered in \cite[Thm. 1]{dabora2023}. In the dual problem, there is no statement in the literature of the capacity for {\em general} non-information-stable channels {\em subject to an average power constraint}, but rather {\em only subject to a per-codeword} constraint\footnote{The given per-codeword power constraint $P$ on a sequence of channel input symbols 
    $\{x_{u}[i]\}_{i=0}^{l}$ with a blocklength of $l$ is expressed as
    \vspace{-0.2cm}
    \begin{equation*}
        \frac{1}{l}\sum_{i=0}^{l-1} (x_{u}[i])^{2}\leq P, \quad u\in\malp{U},
    \end{equation*} 

    \vspace{-0.2cm}
    \noindent where $u$ and $\malp{U}$ are the index and the set of  messages, respectively. See \cite[Eqn.~(6)]{dabora2023}.}.
    Then, in order to meet the upper bound in  \cite[Thm. 1]{dabora2023}, it was  assumed that  the input process $\{X_{in}[i]\}_{i=0}^{l-1} \equiv \mvec{X}_{in,l}$  has to satisfy both the standard trace constraint  at every $l\in\mN^+$, $\frac{1}{l} \cdot \tr\{\mmat{C}_{\mvec{X}_{in,l}}\} \leq P$, as well as {\em an additional asymptotic constraint} $\lim_{l \to \infty}\frac{1}{l^2} \cdot \tr\{(\mmat{C}_{\mvec{X}_{in,l}})^{2}\} = 0$, and subsequently \cite[Thm. 3.6.1]{han2010} was used. The fact that for the source coding problem there exists a characterization for the \gls{rdf} subject to an {\em average distortion constraint}, but with the additional requirement of uniform integrability, shows that these two dual problems are inherently different. 
\end{remark}


\section{Conclusion}   
\label{conclusion}
In this work, we have derived the rate-distortion function for the lossy compression of \gls{dt} \gls{wsacs} Gaussian processes with memory,  obtained via asynchronously sampling \gls{ct} \gls{wscs} Gaussian source processes with a sampling interval shorther than the maximal autocorrelation length of the \gls{ct} source process. 
This problem models  the compression of communications signals, e.g., for relaying or for storage.
We first explained that when sampling a \gls{ct} \gls{wscs} process, synchronous sampling  results in a \gls{dt} \gls{wscs} process while asynchronous sampling results in a \gls{dt} \gls{wsacs} process. In the latter case, the resulting process is non-stationary and non-information-stable, and therefore conventional information-theoretic arguments commonly employed in the characterization of \glspl{rdf} are typically inapplicable. This necessitates carrying out the analysis within the information-spectrum framework. 
Subsequently we derived the \gls{rdf} for  lossy compression of  \gls{dt} \gls{wsacs} random Gaussian processes, assuming initial sampling phase synchronization between the source and the destination and that subsequent sequences are compressed without delay. The relationship with synchronous sampling, as well as the difference between this result and the dual problem studied in \cite[Thm.~1]{dabora2023} were addressed in the discussion. Interesting problems for future work include analysis of the case where delay between codewords is allowed, and the proof of conjecture which facilitates fixing the sampling phase in the  Thm.~\ref{thm: main theorem} and in Corollary \ref{corr:No_delay_fixed rate}.




\appendices   

\setcounter{lemma}{0}
\renewcommand{\thelemma}{\thesection.\arabic{lemma}}
\setcounter{equation}{0}
\renewcommand{\theequation}{\thesection.\arabic{equation}}

\section{Proof of Lemma \ref{Lemma:intergrability}}
\label{appendix:lemma_intefrability}
As derived in \cite[Proof of Lemma A.1]{dabora2023}, we can write $W_l=\frac{1}{l}\sum_{k=0}^{\rank\{\Cxl\}-1}\lambda_{\Xvecl}^{(k)}\Gamma^2_k$, where $\Gamma^2_k\sim\chi^2(1)$ are \gls{iid} central chi-square \glspl{rv} with a single degree-of-freedom, and $\lambda_{\Xvecl}^{(k)}$, $k=0, 1, 2, ..., \rank\{\Cxl\}-1$, are the non-zero eigenvalues of $\Cxl$, thus $\lambda_{\Xvecl}^{(k)}\in\mR^{++}$. Note that $W_l\ge 0$. 
We now obtain $\mE\{W_l\} = \frac{1}{l}\sum_{k=0}^{\rank\{\Cxl\}-1}\lambda_{\Xvecl}^{(k)}\mE\{\Gamma^2_k\} = \frac{1}{l}\sum_{k=0}^{\rank\{\Cxl\}-1}\lambda_{\Xvecl}^{(k)} = \frac{1}{l}\tr\{\Cxl\}\le \beta$, and 
    \begin{align*}
        \mE\{|W_l|^2\} & = \var\big\{W_l\big\}+\big(\mE\{W_l\}\big)^2\\
        & = \frac{1}{l^2}\!\!\!\!\!\sum_{k=0}^{\rank\{\Cxl\}-1}\!\!\!\!\!\!\!\Big(\lambda_{\Xvecl}^{(k)}\Big)^2\!\!\var\{\Gamma^2_k\} + \left(\frac{1}{l}\tr\{\Cxl\}\!\right)^2\\
        & = \frac{2}{l^2}\sum_{k=0}^{\rank\{\Cxl\}-1}\Big(\lambda_{\Xvecl}^{(k)}\Big)^2 + \left(\frac{1}{l}\tr\{\Cxl\}\right)^2\\
        & \stackrel{(a)}{\le} 2\cdot\left(\frac{1}{l} \sum_{k=0}^{\rank\{\Cxl\}-1}\!\!\!\!\lambda_{\Xvecl}^{(k)}\right)^2\!\! + \left(\frac{1}{l}\tr\{\Cxl\}\right)^2\\
        & \le 3\cdot\beta^2 <\infty,
    \end{align*}
    where (a) follows as $\lambda_{\Xvecl}^{(k)}> 0$. It therefore follows that $\mathop{\sup}\limits_{l\in\mN^+}\mE\big\{\big|W_l\big|^2\big\}<\infty$, hence, by \cite[Eqn. (3.18)]{billingsley1999convergence}, $\frac{1}{l}\big(\Xvecl\big)^T\cdot \Xvecl$, $l\in\mN^+$, defined in the lemma, is uniformly integrable.


\section{Proof of Thm.~\ref{thm: main theorem}}
\label{Appendix:proof_of_main_theorem}

\vspace{-0.2cm}

We start with the converse part. First, consider the following lemma, which states the converse for a given sampling phase:

\begin{lemma}[Converse for a given sampling phase $\phi_s$]
\label{lem: converse for given sampling phase}
Consider the scenario in Thm.~\ref{thm: main theorem}, when  
the sampling phase of every source message,  $\phi_s\in[0,T_c)$, is fixed and given; then, for a given average distortion constraint $D$, the \gls{rdf} $R_{\epsilon}^{\phi_s}(D)$ satisfies 

\vspace{-0.6cm}
\[
    R_{\epsilon}^{\phi_s}(D) \geq \
     \limsup_{l\to\infty} \inf_{\mmat{C}_{\mvec{S}_{\epsilon,l}^{\phi_s}}\in \mC_{\mvec{S}_{\epsilon,l}^{\phi_s}}}  \frac{1}{2l} \log\bigg(\frac{\det(\mmat{C}_{\mvec{X}_{\epsilon,l}^{\phi_s}})}{\det(\mmat{C}_{\mvec{S}_{\epsilon,l}^{\phi_s}})}\bigg).
\]
\vspace{-0.3cm}
\end{lemma}
\begin{IEEEproof}
We show that for any sequence of lossy source codes $(M(l,D),l, D)$ asymptotically satisfying the average distortion constraint $D$, 
the code rate 
$R(l,D)=\frac{1}{l}\log M(l,D)$ satisfies 
\[ 
    \limsup_{l\to\infty}R(l,D) \geq \limsup_{l\to\infty} \inf_{\substack{p(\mr{\mvec{X}}_{\epsilon,l}^{\phi_s} | \mvec{X}_{\epsilon,l}^{\phi_s}):\\ \mE\{\maa{d}_{se}(\mvec{X}_{\epsilon,l}^{\phi_s}, \mr{\mvec{X}}_{\epsilon,l}^{\phi_s})\} \leq D}} \frac{1}{l} I(\mvec{X}_{\epsilon,l}^{\phi_s};\mr{\mvec{X}}_{\epsilon,l}^{\phi_s}).
\]

Consider a sequence of  lossy source codes $(\tilde{M}(l,D),l,D)$ defined by an encoder $\tilde{f}_l(\cdot)$ and a decoder $\tilde{g}_l(\cdot)$, such that 
for asymptotically long blocklengths $l$, the given average distortion constraint $D$ is satisfied, namely,
\vspace{-0.1cm}
\begin{equation}
\label{definition of given average distortion constraint}
\limsup_{l\to\infty} \mE\{\maa{d}_{se}(\mvec{X}_{\epsilon,l}^{\phi_s}, \mr{\tilde{\mvec{X}}}_{\epsilon,l}^{\phi_s})\} \leq D,
\end{equation}

\vspace{-0.2cm}
where $\mr{\tilde{\mvec{X}}}_{\epsilon,l}^{\phi_s}=\tilde{g}_l\big(\tilde{f}_l(\mvec{X}_{\epsilon,l}^{\phi_s})\big)$.
Let $\mE\{\maa{d}_{se}(\mvec{X}_{\epsilon,l}^{\phi_s}, \mr{\tilde{\mvec{X}}}_{\epsilon,l}^{\phi_s})\}$ be the distortion achieved by the encode-decoder pair.
Applying \cite[Thm.~2.8]{kostina2013} 
we obtain that for a finite fixed blocklength $l$ it holds that
\begin{multline}
    \log{\tilde{M}(l,D)} \geq \!\!\!\!\!\!\!\!
    \inf_{\substack{p(\mr{\mvec{X}}_{\epsilon,l}^{\phi_s} | \mvec{X}_{\epsilon,l}^{\phi_s}): \\ \mE\{\maa{d}_{se}(\mvec{X}_{\epsilon,l}^{\phi_s}, \mr{\mvec{X}}_{\epsilon,l}^{\phi_s})\} \leq \mE\{\maa{d}_{se}(\mvec{X}_{\epsilon,l}^{\phi_s}, \mr{\tilde{\mvec{X}}}_{\epsilon,l}^{\phi_s})\}}} \!\!\!\!\!\!\!\!\!\!\!\!I(\mvec{X}_{\epsilon,l}^{\phi_s};\mr{\mvec{X}}_{\epsilon,l}^{\phi_s}).\notag
\end{multline}

Equivalently, the code rate $\tR(l,D)$ for code $(\tilde{M}(l,D),l,D)$ satisfies 
\begin{align*}
    \!\!\!\tR(l,D) & 
    \geq \!\!\!\!\!\!\inf_{\substack{p(\mr{\mvec{X}}_{\epsilon,l}^{\phi_s} | \mvec{X}_{\epsilon,l}^{\phi_s}): \\\mE\{\maa{d}_{se}(\mvec{X}_{\epsilon,l}^{\phi_s}, \mr{\mvec{X}}_{\epsilon,l}^{\phi_s})\} \leq \mE\{\maa{d}_{se}(\mvec{X}_{\epsilon,l}^{\phi_s}, \mr{\tilde{\mvec{X}}}_{\epsilon,l}^{\phi_s})\}}} \!\!\!\frac{1}{l} I(\mvec{X}_{\epsilon,l}^{\phi_s};\mr{\mvec{X}}_{\epsilon,l}^{\phi_s}).
\end{align*}
%
As the inequality holds for \emph{every $l$}, then taking limit superior w.r.t. the blocklength $l$ we obtain
\begin{multline}
    \label{limsup_l_eqn}
    \!\!\!\!\!\limsup_{l\to\infty}\tR(l,D) \\
    \geq \limsup_{l\to\infty} \!\!\!\!\!\!\!\inf_{\substack{p(\mr{\mvec{X}}_{\epsilon,l}^{\phi_s} | \mvec{X}_{\epsilon,l}^{\phi_s}):\\ \mE\{\maa{d}_{se}(\mvec{X}_{\epsilon,l}^{\phi_s}, \mr{\mvec{X}}_{\epsilon,l}^{\phi_s})\} \leq \mE\{\maa{d}_{se}(\mvec{X}_{\epsilon,l}^{\phi_s}, \mr{\tilde{\mvec{X}}}_{\epsilon,l}^{\phi_s})\}}}\!\!\!\!\!\!\!
    \frac{1}{l} I(\mvec{X}_{\epsilon,l}^{\phi_s};\mr{\mvec{X}}_{\epsilon,l}^{\phi_s}).
\end{multline}

By \cite[Thm.~4.1.9.(a)]{trench2022}, for any $\delta\in\mRdplus$, there exists an associated blocklength $l_{\delta}^{1}\in\mNplus$, s.t. for any blocklength $l\geq l_{\delta}^{1}$:
\begin{align*}
    \mE\{\maa{d}_{se}(\mvec{X}_{\epsilon,l}^{\phi_s}, \mr{\tilde{\mvec{X}}}_{\epsilon,l}^{\phi_s})\} & < \limsup_{l\to\infty} \mE\{\maa{d}_{se}(\mvec{X}_{\epsilon,l}^{\phi_s}, \mr{\tilde{\mvec{X}}}_{\epsilon,l}^{\phi_s})\} + \delta \nonumber\\
    & 
    \labelrel\leq{step b 1014pm} D + \delta,
\end{align*}
where \eqref{step b 1014pm} follows from Eqn.~\eqref{definition of given average distortion constraint}.
Therefore, when $\delta\to 0$ and the blocklength $l$ is sufficiently large, we asymptotically obtain
\begin{equation}
\label{asymptotical distortion}
\mE\{\maa{d}_{se}(\mvec{X}_{\epsilon,l}^{\phi_s}, \mr{\tilde{\mvec{X}}}_{\epsilon,l}^{\phi_s})\} \leq D.
\end{equation}

When substituting Eqn.~\eqref{asymptotical distortion} back into Eqn.~\eqref{limsup_l_eqn}, the average distortion constraint for minimizing the objective function $\frac{1}{l} I(\mvec{X}_{\epsilon,l}^{\phi_s};\mr{\mvec{X}}_{\epsilon,l}^{\phi_s})$ is relaxed, 
thus, the infimum of the objective function $\frac{1}{l} I(\mvec{X}_{\epsilon,l}^{\phi_s};\mr{\mvec{X}}_{\epsilon,l}^{\phi_s})$ under this looser  constraint is not larger than the original infimum, i.e., 
\begin{align*}
    \!\!\!\!\!&\!\!\!\!\!\limsup_{l\to\infty}\tR(l,D) \notag\\
    & \geq \limsup_{l\to\infty} \!\!\!\!\! \!\!\!\inf_{\substack{p(\mr{\mvec{X}}_{\epsilon,l}^{\phi_s} | \mvec{X}_{\epsilon,l}^{\phi_s}): \\ \mE\{\maa{d}_{se}(\mvec{X}_{\epsilon,l}^{\phi_s}, \mr{\mvec{X}}_{\epsilon,l}^{\phi_s})\} \leq \mE\{\maa{d}_{se}(\mvec{X}_{\epsilon,l}^{\phi_s}, \mr{\tilde{\mvec{X}}}_{\epsilon,l}^{\phi_s})\}}} \!\!\!\!\!\!\!\!
    \frac{1}{l} I(\mvec{X}_{\epsilon,l}^{\phi_s};\mr{\mvec{X}}_{\epsilon,l}^{\phi_s})\\
    & \geq \limsup_{l\to\infty} \inf_{\substack{p(\mr{\mvec{X}}_{\epsilon,l}^{\phi_s} | \mvec{X}_{\epsilon,l}^{\phi_s}): \\    \mE\{\maa{d}_{se}(\mvec{X}_{\epsilon,l}^{\phi_s}, \mr{\mvec{X}}_{\epsilon,l}^{\phi_s})\} \leq D}} \frac{1}{l} I(\mvec{X}_{\epsilon,l}^{\phi_s};\mr{\mvec{X}}_{\epsilon,l}^{\phi_s}).
\end{align*}

As this holds for any sequence of codes asymptotically satisfying the 
average distortion constraint $D$, it holds also for the sequence of codes with the asymptotic rate, which is the \gls{rdf} $ R_{\epsilon}^{\phi_s}(D)$, and thus we obtain
\begin{multline}
    \label{eqn:RDF_lower_bound_with_mutual_information}
    R_{\epsilon}^{\phi_s}(D) \geq 
    \limsup_{l\to\infty} \inf_{\substack{p(\mr{\mvec{X}}_{\epsilon,l}^{\phi_s} | \mvec{X}_{\epsilon,l}^{\phi_s}): \\ \mE\{\maa{d}_{se}(\mvec{X}_{\epsilon,l}^{\phi_s}, \mr{\mvec{X}}_{\epsilon,l}^{\phi_s})\} \leq D}}\!\!\! \frac{1}{l} I(\mvec{X}_{\epsilon,l}^{\phi_s};\mr{\mvec{X}}_{\epsilon,l}^{\phi_s}).
\end{multline}

\vspace{-0.2cm}
Next, we show that the vector $\mr{\mvec{X}}_{\epsilon,l}^{\phi_s}$ that attains the infimum in \eqref{eqn:RDF_lower_bound_with_mutual_information} for a fixed $l$ is  Gaussian. 
%
We denote the optimal sequence of reconstruction symbols as $\mr{\mvec{X}}_{\epsilon,l}^{\phi_s,*}$ and define $\mvec{S}_{\epsilon,l}^{\phi_s,*} \triangleq \mvec{X}_{\epsilon,l}^{\phi_s}-\mr{\mvec{X}}_{\epsilon,l}^{\phi_s,*}$. The objective function $\frac{1}{l} I(\mvec{X}_{\epsilon,l}^{\phi_s};\mr{\mvec{X}}_{\epsilon,l}^{\phi_s})$ is lower bounded as follows:
\begin{subequations}
\begin{align}
    \frac{1}{l} I(\mvec{X}_{\epsilon,l}^{\phi_s};\mr{\mvec{X}}_{\epsilon,l}^{\phi_s}) & = \frac{1}{l} h(\mvec{X}_{\epsilon,l}^{\phi_s})-\frac{1}{l} h(\mvec{X}_{\epsilon,l}^{\phi_s} | \mr{\mvec{X}}_{\epsilon,l}^{\phi_s}) \notag \\
    & = \frac{1}{l} h(\mvec{X}_{\epsilon,l}^{\phi_s})-\frac{1}{l} h(\mvec{X}_{\epsilon,l}^{\phi_s} - \mr{\mvec{X}}_{\epsilon,l}^{\phi_s} | \mr{\mvec{X}}_{\epsilon,l}^{\phi_s}) \notag\\
    & \labelrel\geq{step a Der Gauss} \frac{1}{l} h(\mvec{X}_{\epsilon,l}^{\phi_s})-\frac{1}{l} h(\mvec{X}_{\epsilon,l}^{\phi_s} - \mr{\mvec{X}}_{\epsilon,l}^{\phi_s} ) \notag\\
    & = \frac{1}{l} h(\mvec{X}_{\epsilon,l}^{\phi_s})-\frac{1}{l} h(\mvec{S}_{\epsilon,l}^{\phi_s}) \notag \\
    & \labelrel\geq{step a 1021pm} \frac{1}{l} h(\mvec{X}_{\epsilon,l}^{\phi_s})-\frac{1}{l} h\big(\mGd(\zerovec_l, \mmat{C}_{\mvec{S}_{\epsilon,l}^{\phi_s}})\big) \notag\\
    & \label{b 1021pm} = \frac{1}{2l} \log\bigg(\frac{\det(\mmat{C}_{\mvec{X}_{\epsilon,l}^{\phi_s}})}{\det(\mmat{C}_{\mvec{S}_{\epsilon,l}^{\phi_s}})}\bigg) \\
    & = \label{c 0238am} \frac{1}{l} I(\mvec{X}_{\epsilon,l}^{\phi_s};\mr{\mvec{X}}_{\epsilon,l}^{\phi_s,*}),
\end{align}
\end{subequations}
where \eqref{step a Der Gauss} follows as conditioning reduces differential entropy \cite[Ch. 8.6]{cover2006}; and \eqref{step a 1021pm} follows from \cite[Thm.~8.6.5]{cover2006}. Hence, the vector $\mvec{S}_{\epsilon,l}^{\phi_s}$\footnote{Here we note that as $\mmat{C}_{\mvec{X}_{\epsilon,l}^{\phi_s}}\succ 0$ then also $\mmat{C}_{\mvec{S}_{\epsilon,l}^{\phi_s}} \succ 0$, as otherwise the rate is infinite, while a finite rate is readily achievable, e.g., by fixing all the diagonal elements of $\mmat{C}_{\mvec{S}_{\epsilon,l}^{\phi_s}}$ to $D$.}
that minimizes the objective,  $\mvec{S}_{\epsilon,l}^{\phi_s,*}$ is a Gaussian vector that is independent of $\mr{\mvec{X}}_{\epsilon,l}^{\phi_s}$. As $\mr{\mvec{X}}_{\epsilon,l}^{\phi_s,*} = \mvec{X}_{\epsilon,l}^{\phi_s} - \mvec{S}_{\epsilon,l}^{\phi_s,*}$, where $\mvec{X}_{\epsilon,l}^{\phi_s}$ and $\mvec{S}_{\epsilon,l}^{\phi_s,*}$ are both Gaussian vectors, $\mr{\mvec{X}}_{\epsilon,l}^{\phi_s}$ must also be a Gaussian vector \cite[Rmk.~4.5.2]{ahsanullah2014}. Moreover, as $\mr{\mvec{X}}_{\epsilon,l}^{\phi_s}$ and $\mvec{S}_{\epsilon,l}^{\phi_s,*}$ are mutually independent \glspl{rv} then $\mmat{C}_{\mvec{X}_{\epsilon,l}^{\phi_s}} =  \mmat{C}_{\mr{\mvec{X}}_{\epsilon,l}^{\phi_s}} + \mmat{C}_{\mvec{S}_{\epsilon,l}^{\phi_s}}$, where all matrices are positive semidefinite. Hence, it must also hold that  $\mmat{C}_{\mvec{X}_{\epsilon,l}^{\phi_s}} \succeq \mmat{C}_{\mvec{S}_{\epsilon,l}^{\phi_s}}$ 
 and therefore $\mmat{C}_{\mvec{S}_{\epsilon,l}^{\phi_s}}\in \mC_{\mvec{S}_{\epsilon,l}^{\phi_s}}$.

It follows that the lower bound in Lemma \ref{lem: converse for given sampling phase} can be explicitly stated as 
\[
    R_{\epsilon}^{\phi_s}(D) \geq \limsup_{l\to\infty} 
    \inf_{\mmat{C}_{\mvec{S}_{\epsilon,l}^{\phi_s}}\in \mC_{\mvec{S}_{\epsilon,l}^{\phi_s}}}
    \frac{1}{2l} \log\bigg(\frac{\det(\mmat{C}_{\mvec{X}_{\epsilon,l}^{\phi_s}})}{\det(\mmat{C}_{\mvec{S}_{\epsilon,l}^{\phi_s}})}\bigg),
\]
which completes the proof of the lemma.
\end{IEEEproof}
It is emphasized  since $\mmat{C}_{\mvec{X}_{\epsilon,l}^{\phi_s}} \succeq \mmat{C}_{\mvec{S}_{\epsilon,l}^{\phi_s}}$ holds, it follows  that $\log\bigg(\frac{\det(\mmat{C}_{\mvec{X}_{\epsilon,l}^{\phi_s}})}{\det(\mmat{C}_{\mvec{S}_{\epsilon,l}^{\phi_s}})}\bigg)\ge 0 $.
Next, we proceed to the proof of the achievability part, stated below:

\begin{lemma}[Achievability for a given sampling phase $\phi_s$]
\label{lem: achievability for given sampling phase}
For the scenario in Thm.~\ref{thm: main theorem}, when 
the sampling phase of every source message, $\phi_s\in[0,T_c)$, is fixed and given, we have that, for a given average distortion constraint $D$, the \gls{rdf} $R_{\epsilon}^{\phi_s}(D)$
satisfies 
\[R_{\epsilon}^{\phi_s}(D) \leq 
    \limsup_{l\to\infty} \inf_{\mmat{C}_{\mvec{S}_{\epsilon,l}^{\phi_s}}\in \mC_{\mvec{S}_{\epsilon,l}^{\phi_s}}}  \frac{1}{2l} \log\bigg(\frac{\det(\mmat{C}_{\mvec{X}_{\epsilon,l}^{\phi_s}})}{\det(\mmat{C}_{\mvec{S}_{\epsilon,l}^{\phi_s}})}\bigg).
\]
\end{lemma}

\begin{IEEEproof}
Let
$F^{\phi_s,l}_{X_{\epsilon},\mr{X}_{\epsilon}^{*}}$ denote the joint \gls{cdf} of the vector $\mvec{X}_{\epsilon,l}^{\phi_s}$ and the (Gaussian) vector 
$\mr{\mvec{X}}_{\epsilon,l}^{\phi_s,*}$ such that 
\[
   \mvec{X}_{\epsilon,l}^{\phi_s}  = \mr{\mvec{X}}_{\epsilon,l}^{\phi_s,*}+\mvec{S}_{\epsilon,l}^{\phi_s,*}, 
\]
where $\mvec{S}_{\epsilon,l}^{\phi_s,*}$ is a Gaussian vector  independent of $\mr{\mvec{X}}_{\epsilon,l}^{\phi_s,*}$, which minimizes  \eqref{b 1021pm} subject to $\frac{1}{l}\tr\left\{\mmat{C}_{\mvec{S}_{\epsilon,l}^{\phi_s}}\right\} \leq D$ and to $\mmat{C}_{\mvec{X}_{\epsilon,l}^{\phi_s}} \succeq \mmat{C}_{\mvec{S}_{\epsilon,l}^{\phi_s}}$ (i.e., $\mmat{C}_{\mvec{S}_{\epsilon,l}^{\phi_s}}\in\mC_{\mvec{S}_{\epsilon,l}^{\phi_s}})$. 
Let $Z_{l}(F^{\phi_s}_{X_{\epsilon},\mr{X}_{\epsilon}^{*}})$ denote the mutual information density rate (see \cite[Eqn.~(1.5)]{verdu1994} and \cite[Def.~3.2.1]{han2010}) between $\mvec{X}_{\epsilon,l}^{\phi_s}$ and 
$\mr{\mvec{X}}_{\epsilon,l}^{\phi_s,*}$:
\begin{equation*}
    Z_{l}(F^{\phi_s}_{X_{\epsilon},\mr{X}_{\epsilon}^{*}}) \triangleq \frac{1}{l} \log{\bigg(\frac{p_{X_{\epsilon,l}^{\phi_s} | \mr{X}_{\epsilon,l}^{\phi_s,*}}(\mvec{X}_{\epsilon,l}^{\phi_s} | \mr{\mvec{X}}_{\epsilon,l}^{\phi_s,*})}{p_{X_{\epsilon,l}^{\phi_s}}(\mvec{X}_{\epsilon,l}^{\phi_s})}\bigg)}.
\end{equation*}
The mutual information density $V_{\epsilon,l}^{\phi_s} \triangleq l \cdot Z_{l}(F^{\phi_s}_{X_{\epsilon},\mr{X}_{\epsilon}^{*}})$ can be expressed as
\begin{align*}
    V_{\epsilon,l}^{\phi_s} & = \log{\Big(\frac{p_{X_{\epsilon,l}^{\phi_s} | \mr{X}_{\epsilon,l}^{\phi_s,*}}(\mvec{X}_{\epsilon,l}^{\phi_s} | \mr{\mvec{X}}_{\epsilon,l}^{\phi_s,*})}{p_{X_{\epsilon,l}^{\phi_s}}(\mvec{X}_{\epsilon,l}^{\phi_s})}\Big)} \\
    & \labelrel\eqdist{step z 0603pm} \log{\big(p_{S_{\epsilon,l}^{\phi_s,*}}(\mvec{S}_{\epsilon,l}^{\phi_s,*})\big)} - \log{\big(p_{X_{\epsilon,l}^{\phi_s}}(\mvec{X}_{\epsilon,l}^{\phi_s})\big)} \\
    & \labelrel={step a 1243am} \frac{1}{2} \log\bigg(\frac{\det(\mmat{C}_{\mvec{X}_{\epsilon,l}^{\phi_s}})}{\det(\mmat{C}_{\mvec{S}_{\epsilon,l}^{\phi_s,*}})}\bigg) \\ 
    & \quad + \frac{1}{2}\log{(e)} \big((\mvec{X}_{\epsilon,l}^{\phi_s})^{T}(\mmat{C}_{\mvec{X}_{\epsilon,l}^{\phi_s}})^{-1}\mvec{X}_{\epsilon,l}^{\phi_s}\big) \\
    & \quad - \frac{1}{2}\log{(e)} \big((\mvec{S}_{\epsilon,l}^{\phi_s,*})^{T}(\mmat{C}_{\mvec{S}_{\epsilon,l}^{\phi_s,*}})^{-1}\mvec{S}_{\epsilon,l}^{\phi_s,*}\big),
\end{align*}
where \eqref{step z 0603pm} follows from the statistical independence of the vectors $\mvec{S}_{\epsilon,l}^{\phi_s,*}$ and $\mr{\mvec{X}}_{\epsilon,l}^{\phi_s,*}$ (see \cite[Sec.~10.3.2]{cover2006}); and \eqref{step a 1243am} follows from the Gaussianity of the vectors $\mvec{X}_{\epsilon,l}^{\phi_s}$ and $\mvec{S}_{\epsilon,l}^{\phi_s,*}$. Next, define the \gls{rv} $\tilde{V}_{\epsilon,l}^{\phi_s}$ such that $V_{\epsilon,l}^{\phi_s} = \frac{1}{2} \log\bigg(\frac{\det(\mmat{C}_{\mvec{X}_{\epsilon,l}^{\phi_s}})}{\det(\mmat{C}_{\mvec{S}_{\epsilon,l}^{\phi_s,*}})}\bigg) + \frac{1}{2}\log(e) \cdot\tilde{V}_{\epsilon,l}^{\phi_s}$:
\begin{align}
    & \nonumber \tilde{V}_{\epsilon,l}^{\phi_s}  \triangleq (\mvec{X}_{\epsilon,l}^{\phi_s})^{T}(\mmat{C}_{\mvec{X}_{\epsilon,l}^{\phi_s}})^{-1}\mvec{X}_{\epsilon,l}^{\phi_s} - (\mvec{S}_{\epsilon,l}^{\phi_s,*})^{T}(\mmat{C}_{\mvec{S}_{\epsilon,l}^{\phi_s,*}})^{-1}\mvec{S}_{\epsilon,l}^{\phi_s,*} \\
    \begin{split}
    & = \begin{bmatrix} \mvec{X}_{\epsilon,l}^{\phi_s} \\ {\mvec{S}}_{\epsilon,l}^{\phi_s,*} \end{bmatrix}^{T}  \cdot 
    \begin{bmatrix} 
        (\mmat{C}_{\mvec{X}_{\epsilon,l}^{\phi_s}})^{-1}  & \zeromat_{l} \\ 
        \zeromat_{l} &  -(\mmat{C}_{\mvec{S}_{\epsilon,l}^{\phi_s}})^{-1}
    \end{bmatrix} \cdot  \begin{bmatrix} \mvec{X}_{\epsilon,l}^{\phi_s} \\ {\mvec{S}}_{\epsilon,l}^{\phi_s,*} \end{bmatrix}.\notag
    \end{split}
\end{align}

We also define the matrix $\tilde{\mmat{C}}_{\epsilon,l}^{\phi_s}$ as follows:
\begin{equation*}
    \tilde{\mmat{C}}_{\epsilon,l}^{\phi_s} \triangleq 
    \begin{bmatrix} 
        (\mmat{C}_{\mvec{X}_{\epsilon,l}^{\phi_s}})^{-1}  & \zeromat_{l} \\ 
        \zeromat_{l} &  -(\mmat{C}_{\mvec{S}_{\epsilon,l}^{\phi_s}})^{-1}
    \end{bmatrix},
\end{equation*}
 and 
write $\tilde{V}_{\epsilon,l}^{\phi_s} = \begin{bmatrix} \mvec{X}_{\epsilon,l}^{\phi_s} \\ {\mvec{S}}_{\epsilon,l}^{\phi_s,*} \end{bmatrix}^{T} \tilde{\mmat{C}}_{\epsilon,l}^{\phi_s} \begin{bmatrix} \mvec{X}_{\epsilon,l}^{\phi_s} \\ {\mvec{S}}_{\epsilon,l}^{\phi_s,*}
\end{bmatrix}$.

Next consider the autocorrelation matrix $\mmat{C}_{\mvec{X}_{\epsilon,l}^{\phi_s}, {\mvec{S}}_{\epsilon,l}^{\phi_s,*}} \triangleq \mE\Big\{\begin{bmatrix}(\mvec{X}_{\epsilon,l}^{\phi_s})^{T} & ({\mvec{S}}_{\epsilon,l}^{\phi_s,*})^{T}\end{bmatrix}^{T} \cdot \begin{bmatrix}(\mvec{X}_{\epsilon,l}^{\phi_s})^{T} & ({\mvec{S}}_{\epsilon,l}^{\phi_s,*})^{T}\end{bmatrix}\Big\}= 
\begin{bmatrix} \mmat{C}_{\mvec{X}_{\epsilon,l}^{\phi_s}} & \mmat{C}_{\mvec{X}_{\epsilon,l}^{\phi_s}  {\mvec{S}}_{\epsilon,l}^{\phi_s,*}} \\ \mmat{C}_{{\mvec{S}}_{\epsilon,l}^{\phi_s,*}  \mvec{X}_{\epsilon,l}^{\phi_s}} & \mmat{C}_{{\mvec{S}}_{\epsilon,l}^{\phi_s,*}} \end{bmatrix}$, 
where 
$\mmat{C}_{\mvec{X}_{\epsilon,l}^{\phi_s} {\mvec{S}}_{\epsilon,l}^{\phi_s,*}}\triangleq \mE\{\mvec{X}_{\epsilon,l}^{\phi_s}\cdot ({\mvec{S}}_{\epsilon,l}^{\phi_s,*})^{T}\}$ and $\mmat{C}_{{\mvec{S}}_{\epsilon,l}^{\phi_s,*}\mvec{X}_{\epsilon,l}^{\phi_s}}\triangleq \mE\{{\mvec{S}}_{\epsilon,l}^{\phi_s,*}\cdot (\mvec{X}_{\epsilon,l}^{\phi_s})^{T}\}$. As $\mmat{C}_{\mvec{X}_{\epsilon,l}^{\phi_s}, {\mvec{S}}_{\epsilon,l}^{\phi_s,*}}$ 
 may not be full-rank,
 let us denote 
 $\rank(\mmat{C}_{\mvec{X}_{\epsilon,l}^{\phi_s}, {\mvec{S}}_{\epsilon,l}^{\phi_s,*}}) = 2l-\tilde{l}_{\epsilon,l}$, 
 $\tilde{l}_{\epsilon,l}\in \mN$. To separate the degenerate elements and the non-degenerate elements in the vector $\begin{bmatrix} (\mvec{X}_{\epsilon,l}^{\phi_s})^{T} & ({\mvec{S}}_{\epsilon,l}^{\phi_s,*})^{T} \end{bmatrix}^{T}$, the matrix $\mmat{C}_{\mvec{X}_{\epsilon,l}^{\phi_s},  {\mvec{S}}_{\epsilon,l}^{\phi_s,*}}$ can be decomposed as \cite[Eqn.~(10)]{baktash2017}
\begin{multline}
    \label{a 0111am}
    \mmat{C}_{\mvec{X}_{\epsilon,l}^{\phi_s}, {\mvec{S}}_{\epsilon,l}^{\phi_s,*}} = \begin{bmatrix} \mmat{P}_{\epsilon,l}^{\phi_s,R} & \mmat{P}_{\epsilon,l}^{\phi_s,N} \end{bmatrix} \\ \cdot \begin{bmatrix} \tilde{\mmat{C}}_{\mvec{X}_{\epsilon,l}^{\phi_s} {\mvec{S}}_{\epsilon,l}^{\phi_s,*}} & \zeromat_{(2l-\tilde{l}_{\epsilon,l})\times \tilde{l}_{\epsilon,l}} \\ \zeromat_{\tilde{l}_{\epsilon,l}\times (2l-\tilde{l}_{\epsilon,l})} & \zeromat_{\tilde{l}_{\epsilon,l} \times \tilde{l}_{\epsilon,l}} \end{bmatrix} \begin{bmatrix} \mmat{P}_{\epsilon,l}^{\phi_s,R} & \mmat{P}_{\epsilon,l}^{\phi_s,N} \end{bmatrix}^{T},
\end{multline}
where the $(2l-\tilde{l}_{\epsilon,l})\times (2l-\tilde{l}_{\epsilon,l})$ square matrix $\tilde{\mmat{C}}_{\mvec{X}_{\epsilon,l}^{\phi_s} {{\mvec{S}}}_{\epsilon,l}^{\phi_s,*}}$ is a diagonal matrix having $2l-\tilde{l}_{\epsilon,l}$ non-zero eigenvalues and is thus symmetric and positive definite, the columns of the $2l\times (2l-\tilde{l}_{\epsilon,l})$ matrix $\mmat{P}_{\epsilon,l}^{\phi_s,R}$ form an orthonormal basis of $\range(\tilde{\mmat{C}}_{\mvec{X}_{\epsilon,l}^{\phi_s} {{\mvec{S}}}_{\epsilon,l}^{\phi_s,*}})$, the columns of the $2l\times \tilde{l}_{\epsilon,l}$ matrix $\mmat{P}_{\epsilon,l}^{\phi_s,N}$ form an orthonormal basis of $\nulspc(\tilde{\mmat{C}}_{\mvec{X}_{\epsilon,l}^{\phi_s} {{\mvec{S}}}_{\epsilon,l}^{\phi_s,*}})$, and the $2l\times 2l$ square matrix $\begin{bmatrix} \mmat{P}_{\epsilon,l}^{\phi_s,R} & \mmat{P}_{\epsilon,l}^{\phi_s,N} \end{bmatrix}$ is orthogonal. From Eqn.~\eqref{a 0111am}, we obtain
\begin{equation}
\label{a 0136am}
    \begin{bmatrix} \mmat{P}_{\epsilon,l}^{\phi_s,R} & \mmat{P}_{\epsilon,l}^{\phi_s,N} \end{bmatrix}^{T} \begin{bmatrix} \mvec{X}_{\epsilon,l}^{\phi_s} \\ {\mvec{S}}_{\epsilon,l}^{\phi_s,*} \end{bmatrix} \eqdist \begin{bmatrix} \mmat{B}_{\epsilon,2l- \tilde{l}_{\epsilon,l}}^{\phi_s} \\ \zeromat_{\tilde{l}_{\epsilon,l}\times 1} \end{bmatrix},
\end{equation}
where
\begin{equation}
\label{a 0604pm}
    \mmat{B}_{\epsilon,2l- \tilde{l}_{\epsilon,l}}^{\phi_s}\sim \mGd\Big(\zerovec_{2l- \tilde{l}_{\epsilon,l}}, \tilde{\mmat{C}}_{\mvec{X}_{\epsilon,l}^{\phi_s}{{\mvec{S}}}_{\epsilon,l}^{\phi_s,*}}\Big).
\end{equation}

Using Eqns.~\eqref{a 0136am} and \eqref{a 0604pm} we obtain
\begin{align}
\begin{split}
    \tilde{V}_{\epsilon,l}^{\phi_s} & = \begin{bmatrix} \mvec{X}_{\epsilon,l}^{\phi_s} \\ {\mvec{S}}_{\epsilon,l}^{\phi_s,*} \end{bmatrix}^{T} \cdot \begin{bmatrix} \mmat{P}_{\epsilon,l}^{\phi_s,R} & \mmat{P}_{\epsilon,l}^{\phi_s,N} \end{bmatrix} \begin{bmatrix} \mmat{P}_{\epsilon,l}^{\phi_s,R} & \mmat{P}_{\epsilon,l}^{\phi_s,N} \end{bmatrix}^{T} \\
    &\;\;\;\;\;\; \cdot \tilde{\mmat{C}}_{\epsilon,l}^{\phi_s} \cdot \begin{bmatrix} \mmat{P}_{\epsilon,l}^{\phi_s,R} & \mmat{P}_{\epsilon,l}^{\phi_s,N} \end{bmatrix} \begin{bmatrix} \mmat{P}_{\epsilon,l}^{\phi_s,R} & \mmat{P}_{\epsilon,l}^{\phi_s,N} \end{bmatrix}^{T} \cdot \begin{bmatrix} \mvec{X}_{\epsilon,l}^{\phi_s} \\ \mr{\mvec{X}}_{\epsilon,l}^{\phi_s,*}
    \end{bmatrix}\notag
    \end{split} \\
    & \label{a 0937pm} \eqdist (\mmat{B}_{\epsilon,2l- \tilde{l}_{\epsilon,l}}^{\phi_s})^{T} \cdot (\mmat{P}_{\epsilon,l}^{\phi_s,R})^{T} \cdot \tilde{\mmat{C}}_{\epsilon,l}^{\phi_s} \cdot \mmat{P}_{\epsilon,l}^{\phi_s,R} \cdot \mmat{B}_{\epsilon,2l- \tilde{l}_{\epsilon,l}}^{\phi_s},
\end{align}
where the matrix $(\mmat{P}_{\epsilon,l}^{\phi_s,R})^{T} \cdot \tilde{\mmat{C}}_{\epsilon,l}^{\phi_s} \cdot \mmat{P}_{\epsilon,l}^{\phi_s,R}$ is full-rank, as both $\mmat{P}_{\epsilon,l}^{\phi_s,R}$ and $\tilde{\mmat{C}}_{\epsilon,l}^{\phi_s}$ are full-rank matrices. Since the matrix $\tilde{\mmat{C}}_{\mvec{X}_{\epsilon,l}^{\phi_s} {{\mvec{S}}}_{\epsilon,l}^{\phi_s,*}}$ is symmetric and positive definite, there exists a unique $(2l- \tilde{l}_{\epsilon,l})\times (2l- \tilde{l}_{\epsilon,l})$ square symmetric positive definite matrix $\mmat{R}_{\epsilon,l}^{\phi_s}$ that satisfies $\tilde{\mmat{C}}_{\mvec{X}_{\epsilon,l}^{\phi_s} {{\mvec{S}}}_{\epsilon,l}^{\phi_s,*}} = (\mmat{R}_{\epsilon,l}^{\phi_s})^{2}$ \cite[Sec.~1.1]{bhatia2007}, \cite[Thm.~7.2.6.(a)]{horn2012}. Using $\mmat{R}_{\epsilon,l}^{\phi_s}$ we can express Eqn.~\eqref{a 0937pm} as
\begin{multline}
    \label{a 0953pm}
    \tilde{V}_{\epsilon,l}^{\phi_s} \eqdist (\mmat{B}_{\epsilon,2l- \tilde{l}_{\epsilon,l}}^{\phi_s})^{T} \cdot (\mmat{R}_{\epsilon,l}^{\phi_s})^{-1} \cdot \mmat{R}_{\epsilon,l}^{\phi_s} \cdot (\mmat{P}_{\epsilon,l}^{\phi_s,R})^{T} \cdot \tilde{\mmat{C}}_{\epsilon,l}^{\phi_s} \\
    \cdot \mmat{P}_{\epsilon,l}^{\phi_s,R} \cdot \mmat{R}_{\epsilon,l}^{\phi_s} \cdot (\mmat{R}_{\epsilon,l}^{\phi_s})^{-1} \cdot \mmat{B}_{\epsilon,2l- \tilde{l}_{\epsilon,l}}^{\phi_s}.
\end{multline}
As both $\mmat{R}_{\epsilon,l}^{\phi_s}$ and $(\mmat{P}_{\epsilon,l}^{\phi_s,R})^{T} \cdot \tilde{\mmat{C}}_{\epsilon,l}^{\phi_s} \cdot \mmat{P}_{\epsilon,l}^{\phi_s,R}$ are full-rank matrices, the $(2l- \tilde{l}_{\epsilon,l})\times (2l- \tilde{l}_{\epsilon,l})$ square matrix $\mmat{R}_{\epsilon,l}^{\phi_s} \cdot (\mmat{P}_{\epsilon,l}^{\phi_s,R})^{T} \cdot \tilde{\mmat{C}}_{\epsilon,l}^{\phi_s} \cdot \mmat{P}_{\epsilon,l}^{\phi_s,R} \cdot \mmat{R}_{\epsilon,l}^{\phi_s}$ is also full-rank and thus has $2l-\tilde{l}_{\epsilon,l}$ non-zero eigenvalues. Due to its symmetry, by the spectral decomposition \cite[Sec.~3.8.10.2]{gentle2017} we can write
\begin{subequations}
\label{a 1027pm}
\begin{align}
    \mdtilde{\mmat{C}}_{\epsilon,l}^{\phi_s} & \label{a 1037pm} \triangleq \mmat{R}_{\epsilon,l}^{\phi_s} \cdot (\mmat{P}_{\epsilon,l}^{\phi_s,R})^{T} \cdot \tilde{\mmat{C}}_{\epsilon,l}^{\phi_s} \cdot \mmat{P}_{\epsilon,l}^{\phi_s,R} \cdot \mmat{R}_{\epsilon,l}^{\phi_s} \\
    & \label{b 1036pm} = \mmat{U}_{\epsilon,l}^{\phi_s} \cdot \mmat{\Lambda}_{\epsilon,l}^{\phi_s} \cdot (\mmat{U}_{\epsilon,l}^{\phi_s})^{T},
\end{align}
\end{subequations}
where the columns of the orthogonal matrix $\mmat{U}_{\epsilon,l}^{\phi_s}$ are the normalized eigenvectors of the matrix $\mdtilde{\mmat{C}}_{\epsilon,l}^{\phi_s}$ and the diagonal elements of the diagonal matrix $\mmat{\Lambda}_{\epsilon,l}^{\phi_s}$ are the corresponding eigenvalues of the matrix $\mdtilde{\mmat{C}}_{\epsilon,l}^{\phi_s}$.

Next, define the \gls{rv} $\Gamma_{\epsilon,2l-\tilde{l}_{\epsilon,l}}^{\phi_s}$ as follows:
\begin{equation}
    \label{a 1028pm}
    \Gamma_{\epsilon,2l-\tilde{l}_{\epsilon,l}}^{\phi_s}\triangleq (\mmat{R}_{\epsilon,l}^{\phi_s})^{-1} \cdot \mmat{B}_{\epsilon,2l- \tilde{l}_{\epsilon,l}}^{\phi_s} \sim \mGd(\zerovec_{2l-\tilde{l}_{\epsilon,l}},\idmat_{2l-\tilde{l}_{\epsilon,l}}),
\end{equation}
where the distribution of $\Gamma_{\epsilon,2l-\tilde{l}_{\epsilon,l}}^{\phi_s}$ follows from Eqn.~\eqref{a 0604pm}, and finally
define a vector $\tilde{\Gamma}_{\epsilon,2l-\tilde{l}_{\epsilon,l}}^{\phi_s}$:
\begin{equation}
\label{a 0803pm}
\tilde{\Gamma}_{\epsilon,2l-\tilde{l}_{\epsilon,l}}^{\phi_s}\triangleq (\mmat{U}_{\epsilon,l}^{\phi_s})^{T} \cdot \Gamma_{\epsilon,2l-\tilde{l}_{\epsilon,l}}^{\phi_s} \sim \mGd(\zerovec_{2l-\tilde{l}_{\epsilon,l}},\idmat_{2l-\tilde{l}_{\epsilon,l}}).
\end{equation}

Applying \eqref{a 1027pm}, \eqref{a 1028pm} and \eqref{a 0803pm}  to Eqn.~\eqref{a 0953pm}, we obtain
\begin{align*}
    \tilde{V}_{\epsilon,l}^{\phi_s} & \eqdist \big((\Gamma_{\epsilon,2l-\tilde{l}_{\epsilon,l}}^{\phi_s})^{T} \cdot \mmat{U}_{\epsilon,l}^{\phi_s}\big) \cdot \mmat{\Lambda}_{\epsilon,l}^{\phi_s} \cdot \big((\mmat{U}_{\epsilon,l}^{\phi_s})^{T} \cdot \Gamma_{\epsilon,2l-\tilde{l}_{\epsilon,l}}^{\phi_s}\big) \\
    & \eqdist (\tilde{\Gamma}_{\epsilon,2l-\tilde{l}_{\epsilon,l}}^{\phi_s})^{T} \cdot \mmat{\Lambda}_{\epsilon,l}^{\phi_s} \cdot \tilde{\Gamma}_{\epsilon,2l-\tilde{l}_{\epsilon,l}}^{\phi_s} \\
    & = \sum_{i=0}^{2l-\tilde{l}_{\epsilon,l}-1} \lambda_{\epsilon,i}^{\phi_s} \cdot (\tilde{\gamma}_{\epsilon,i}^{\phi_s})^{2},
\end{align*}
where $\lambda_{\epsilon,i}^{\phi_s}\neq 0$, $0\leq i \leq 2l-\tilde{l}_{\epsilon,l}-1$, denotes the $i$-th diagonal element of the diagonal matrix $\mmat{\Lambda}_{\epsilon,l}^{\phi_s}$ and the \gls{rv} $\tilde{\gamma}_{\epsilon,i}^{\phi_s}$, $0\leq i \leq 2l-\tilde{l}_{\epsilon,l}-1$, denotes the $i$-th element of the vector $\tilde{\Gamma}_{\epsilon,2l-\tilde{l}_{\epsilon,l}}^{\phi_s}$, respectively. From Eqn.~\eqref{a 0803pm}, it follows that $\tilde{\gamma}_{\epsilon,i}^{\phi_s}$ are \gls{iid} standard Gaussian \glspl{rv}, therefore $(\tilde{\gamma}_{\epsilon,i}^{\phi_s})^{2}$, $0\leq i \leq 2l-\tilde{l}_{\epsilon,l}-1$, are \gls{iid} central chi-square \glspl{rv} with a single degree-of-freedom \cite[Sec.~3.8.17]{shynk2012}.

In Eqns.~\eqref{a 0854pm} we show that $\mE\{\tilde{V}_{\epsilon,l}^{\phi_s}\}=0$, where
\begin{figure*}
\begin{subequations}
\label{a 0854pm}
\begin{align}
    \mE\{\tilde{V}_{\epsilon,l}^{\phi_s}\} & = \mE\bigg\{\sum_{i=0}^{2l-\tilde{l}_{\epsilon,l}-1} \lambda_{\epsilon,i}^{\phi_s} \cdot (\tilde{\gamma}_{\epsilon,i}^{\phi_s})^{2}\bigg\} \notag\\
    & \labelrel={step a 0918pm} \sum_{i=0}^{2l-\tilde{l}_{\epsilon,l}-1} \lambda_{\epsilon,i}^{\phi_s} \notag\\
    & \labelrel={step b 0941pm} \tr\{\mmat{R}_{\epsilon,l}^{\phi_s} \cdot (\mmat{P}_{\epsilon,l}^{\phi_s,R})^{T} \cdot \tilde{\mmat{C}}_{\epsilon,l}^{\phi_s} \cdot \mmat{P}_{\epsilon,l}^{\phi_s,R} \cdot \mmat{R}_{\epsilon,l}^{\phi_s}\} \notag\\
    & = \tr\{\tilde{\mmat{C}}_{\epsilon,l}^{\phi_s} \cdot \mmat{P}_{\epsilon,l}^{\phi_s,R} \cdot (\mmat{R}_{\epsilon,l}^{\phi_s})^{2} \cdot (\mmat{P}_{\epsilon,l}^{\phi_s,R})^{T}\} \notag\\
    & = \tr\big\{\tilde{\mmat{C}}_{\epsilon,l}^{\phi_s} \cdot \big(\mmat{P}_{\epsilon,l}^{\phi_s,R} \cdot \tilde{\mmat{C}}_{\mvec{X}_{\epsilon,l}^{\phi_s} {{\mvec{S}}}_{\epsilon,l}^{\phi_s,*}} \cdot (\mmat{P}_{\epsilon,l}^{\phi_s,R})^{T}\big)\big\} \notag\\
    & \labelrel={step c 1118pm} \tr\bigg\{\tilde{\mmat{C}}_{\epsilon,l}^{\phi_s} \cdot \begin{bmatrix} \mmat{P}_{\epsilon,l}^{\phi_s,R} & \mmat{P}_{\epsilon,l}^{\phi_s,N} \end{bmatrix} \begin{bmatrix} \tilde{\mmat{C}}_{\mvec{X}_{\epsilon,l}^{\phi_s} {{\mvec{S}}}_{\epsilon,l}^{\phi_s,*}} & \zeromat_{(2l-\tilde{l}_{\epsilon,l})\times \tilde{l}_{\epsilon,l}} \\ \zeromat_{\tilde{l}_{\epsilon,l}\times (2l-\tilde{l}_{\epsilon,l})} & \zeromat_{\tilde{l}_{\epsilon,l} \times \tilde{l}_{\epsilon,l}} \end{bmatrix} \begin{bmatrix} \mmat{P}_{\epsilon,l}^{\phi_s,R} & \mmat{P}_{\epsilon,l}^{\phi_s,N} \end{bmatrix}^{T}\bigg\} \notag \\
    & \label{d 1119pm} \labelrel={step d 1119pm} \tr\Big\{\tilde{\mmat{C}}_{\epsilon,l}^{\phi_s} \cdot \mmat{C}_{\mvec{X}_{\epsilon,l}^{\phi_s}, {\mvec{S}}_{\epsilon,l}^{\phi_s,*}}\Big\} \\
    & \label{g 0822pm} = 0.
\end{align}
\end{subequations}
\hrulefill
\end{figure*}
in the derivation, \eqref{step a 0918pm} follows as $\mE\{(\tilde{\gamma}_{\epsilon,i}^{\phi_s})^{2}\} = 1$; \eqref{step b 0941pm} follows since the sum of the eigenvalues of a square matrix equals to its trace \cite[Sec.~1.3]{zhang2011} and from Eqns.~\eqref{a 1027pm}; \eqref{step c 1118pm} and \eqref{step d 1119pm} follow from Eqn.~\eqref{a 0111am}; equality to zero in the last step follows by substituting the respective matrices explicitly.
Therefore, we have
\begin{align}
    & \nonumber \mE\{Z_{l}(F^{\phi_s}_{X_{\epsilon},\mr{X}_{\epsilon}^{*}})\} \\
    & = \mE\bigg\{\frac{1}{2l} \log\bigg(\frac{\det(\mmat{C}_{\mvec{X}_{\epsilon,l}^{\phi_s}})}{\det(\mmat{C}_{\mvec{S}_{\epsilon,l}^{\phi_s,*}})}\bigg) + \frac{1}{2l}\log(e) \cdot\tilde{V}_{\epsilon,l}^{\phi_s}\bigg\} \notag\\
    & = \frac{1}{2l} \log\bigg(\frac{\det(\mmat{C}_{\mvec{X}_{\epsilon,l}^{\phi_s}})}{\det(\mmat{C}_{\mvec{S}_{\epsilon,l}^{\phi_s,*}})}\bigg) + \frac{1}{2l}\log(e)\cdot \mE\{\tilde{V}_{\epsilon,l}^{\phi_s}\} \notag\\
    & \labelrel={step a 0853pm} \frac{1}{2l} \log\bigg(\frac{\det(\mmat{C}_{\mvec{X}_{\epsilon,l}^{\phi_s}})}{\det(\mmat{C}_{\mvec{S}_{\epsilon,l}^{\phi_s,*}})}\bigg) \notag\\
    & \labelrel={step b 0236am} \frac{1}{l} I(\mvec{X}_{\epsilon,l}^{\phi_s};\mr{\mvec{X}}_{\epsilon,l}^{\phi_s,*}),
    \label{eqn:avg_Z_thm1}
\end{align}
where \eqref{step a 0853pm} follows from Eqn.~\eqref{g 0822pm}; and \eqref{step b 0236am} follows from Eqn.~\eqref{c 0238am}.

Next, consider $\var\{Z_{l}(F^{\phi_s}_{X_{\epsilon},\mr{X}_{\epsilon}^{*}})\}$. We first compute $\var\{\tilde{V}_{\epsilon,l}^{\phi_s}\}$:
\begin{align}
     \var\{\tilde{V}_{\epsilon,l}^{\phi_s}\}  & = \var\bigg\{\sum_{i=0}^{2l-\tilde{l}_{\epsilon,l}-1} \lambda_{\epsilon,i}^{\phi_s} \cdot (\tilde{\gamma}_{\epsilon,i}^{\phi_s})^{2}\bigg\} \notag\\
    & \labelrel={step_iid_in_var} \sum_{i=0}^{2l-\tilde{l}_{\epsilon,l}-1} \var\{\lambda_{\epsilon,i}^{\phi_s} \cdot (\tilde{\gamma}_{\epsilon,i}^{\phi_s})^{2}\} \notag\\
    & \labelrel={step a 1005pm} 2 \cdot \sum_{i=0}^{2l-\tilde{l}_{\epsilon,l}-1} (\lambda_{\epsilon,i}^{\phi_s})^{2} \notag\\
    & \labelrel={step b 1026pm} 2\cdot \tr\big\{(\mdtilde{\mmat{C}}_{\epsilon,l}^{\phi_s})^{2}\big\} \notag\\
    & \labelrel={step c 1145pm} 2\cdot \tr\Big\{\big(\tilde{\mmat{C}}_{\epsilon,l}^{\phi_s} \cdot \mmat{C}_{\mvec{X}_{\epsilon,l}^{\phi_s}, {\mvec{S}}_{\epsilon,l}^{\phi_s,*}}\big)^{2}\Big\} \notag\\
    & \labelrel={step e 0440pm} 4\cdot \Big(l- \tr\Big\{(\mmat{C}_{\mvec{X}_{\epsilon,l}^{\phi_s}})^{-1}\mmat{C}_{\mvec{S}_{\epsilon,l}^{\phi_s,*}}\Big\}\Big) \notag\\
    & \label{f 0744pm} \labelrel<{step f 0744pm} 4l,
\end{align}
where \eqref{step_iid_in_var} follows as $(\tilde{\gamma}_{\epsilon,i}^{\phi_s})^{2}$ are mutually independent; \eqref{step a 1005pm} follows from $\var\{(\tilde{\gamma}_{\epsilon,i}^{\phi_s})^{2}\} = 2$; \eqref{step b 1026pm} follows from Eqn.~\eqref{b 1036pm}; \eqref{step c 1145pm} follows from Eqns.~\eqref{a 1037pm} and \eqref{d 1119pm}; 
\eqref{step e 0440pm} follows from 
the explicit expressions for the matrices;
and \eqref{step f 0744pm} follows from the positive definiteness of the matrices $\mmat{C}_{\mvec{X}_{\epsilon,l}^{\phi_s}}$ and $\mmat{C}_{\mvec{S}_{\epsilon,l}^{\phi_s,*}}$ and the spectral decomposition of these two matrices, which lead to $\tr\{(\mmat{C}_{\mvec{X}_{\epsilon,l}^{\phi_s}})^{-1}\mmat{C}_{\mvec{S}_{\epsilon,l}^{\phi_s,*}}\} > 0$. Then, we obtain
\begin{align}
    & \var\Big\{Z_{l}\big(F^{\phi_s}_{X_{\epsilon},\mr{X}_{\epsilon}^{*}}\big)\Big\} \notag\\
    & = \var\bigg\{\frac{1}{2l} \log\bigg(\frac{\det(\mmat{C}_{\mvec{X}_{\epsilon,l}^{\phi_s}})}{\det(\mmat{C}_{\mvec{S}_{\epsilon,l}^{\phi_s,*}})}\bigg) + \frac{1}{2l}\log(e) \tilde{V}_{\epsilon,l}^{\phi_s}\bigg\} \notag\\
    & = \frac{1}{4l^{2}} \big(\log(e)\big)^{2} \cdot \var\{\tilde{V}_{\epsilon,l}^{\phi_s}\} \notag\\
    & \label{a 0158am} \labelrel<{step a 0158am} \frac{3}{l},
\end{align}
where \eqref{step a 0158am} follows by substituting Eqn.~\eqref{f 0744pm}.
Applying Chebyshev's inequality \cite[Eqn.~(1.58)]{gallager2013} to the \gls{rv} $Z_{l}\big(F^{\phi_s}_{X_{\epsilon},\mr{X}_{\epsilon}^{*}}\big)$ and incorporating Eqns.~\eqref{eqn:avg_Z_thm1} 
and \eqref{a 0158am}, we obtain (see \cite[Appendix A]{dabora2023})
\begin{equation*}
    \pr\Big\{\Big|Z_{l}\big(F^{\phi_s}_{X_{\epsilon},\mr{X}_{\epsilon}^{*}}\big) - \frac{1}{l} I(\mvec{X}_{\epsilon,l}^{\phi_s};\mr{\mvec{X}}_{\epsilon,l}^{\phi_s,*})\Big| \geq \frac{1}{l^{\frac{1}{3}}}\Big\} < \frac{3}{l^{\frac{1}{3}}}.
\end{equation*}
It thus follows that for any $\delta\in \mRdplus$, there exists an associated blocklength $l_{\delta}^{(2)}\in \mNplus$, such that for any blocklength $l\geq l_{\delta}^{(2)}$,
\begin{equation}
    \label{a 1200am}
    \pr\Big\{Z_{l}(F^{\phi_s}_{X_{\epsilon},\mr{X}_{\epsilon}^{*}})\geq \frac{1}{l} I(\mvec{X}_{\epsilon,l}^{\phi_s};\mr{\mvec{X}}_{\epsilon,l}^{\phi_s,*}) + \delta\Big\} < 3\delta.
\end{equation}

By \cite[Thm.~4.1.9.(a)]{trench2022}, we have that for any $\delta\in \mRdplus$, there exists an associated blocklength $l_{\delta}^{(3)}\in \mNplus$, such that for any blocklength $l\geq l_{\delta}^{(3)}$
\begin{equation*}
    \frac{1}{l} I(\mvec{X}_{\epsilon,l}^{\phi_s};\mr{\mvec{X}}_{\epsilon,l}^{\phi_s,*})< \limsup_{l\to\infty} \frac{1}{l} I(\mvec{X}_{\epsilon,l}^{\phi_s};\mr{\mvec{X}}_{\epsilon,l}^{\phi_s,*}) + \delta.
\end{equation*}
Combining this with 
Eqn.~\eqref{a 1200am} and taking the blocklength $l\geq \max\big\{l_{\delta}^{(2)}, l_{\delta}^{(3)}\big\}$, we obtain:
\begin{equation}
    \label{a 1225am}
    \!\!\!\pr\Big\{Z_{l}(F^{\phi_s}_{X_{\epsilon},\mr{X}_{\epsilon}^{*}}) \!>\! \limsup_{l\to\infty} \frac{1}{l} I(\mvec{X}_{\epsilon,l}^{\phi_s};\mr{\mvec{X}}_{\epsilon,l}^{\phi_s,*}) + 2\delta\Big\}\! <\! 3\delta.
\end{equation}
Then, by the definition of the limit superior in probability,  taking $\delta\to 0$ it follows from Eqn.~\eqref{a 1225am} that 
\begin{align}
    & \nonumber \limsupp_{l\to\infty} Z_{l}(F^{\phi_s}_{X_{\epsilon},\mr{X}_{\epsilon}^{*}}) \\ 
    & \triangleq \inf\big\{\alpha\in\mR | \lim_{l\to\infty}\Pr\{ Z_{l}(F^{\phi_s}_{X_{\epsilon},\mr{X}_{\epsilon}^{*}}) > \alpha \} = 0\big\} \notag\\
    & \label{a 1131pm} \leq \limsup_{l\to\infty} \frac{1}{l} I(\mvec{X}_{\epsilon,l}^{\phi_s};\mr{\mvec{X}}_{\epsilon,l}^{\phi_s,*}).
\end{align}

Next consider the distortion. By design of the reconstruction codebook for the achievable scheme it holds that 
\begin{align}
    &\limsup_{l\rightarrow\infty} \frac{1}{l} \cdot \mE\bigg\{\sum_{i=0}^{l-1} \big(X_{\epsilon}^{\phi_s}[i] - \mr{X}_{\epsilon}^{\phi_s}[i]\big)^{2}\bigg\}\notag\\
    & = \limsup_{l\rightarrow\infty} \frac{1}{l} \tr\left\{\mmat{C}_{\mvec{S}_{\epsilon,l}^{\phi_s,*}}\right\}\notag\\
    &\le D.
    \label{Eqn:Dis_Constr_Satisfied}
\end{align}

Lastly, note that by \cite[Thm. 5.5.1]{han2010}, as the \gls{mse} distortion is uniformly integrable by Lemma \ref{Lemma:intergrability}, and satisfies \eqref{Eqn:Dis_Constr_Satisfied}, then $ \limsupp_{l\to\infty} Z_{l}\big(F^{\phi_s}_{X_{\epsilon},\mr{X}_{\epsilon}^{*}}\big)$ is achievable, and thus, by \eqref{a 1131pm}, $\limsup_{l\to\infty} \frac{1}{l} I(\mvec{X}_{\epsilon,l}^{\phi_s};\mr{\mvec{X}}_{\epsilon,l}^{\phi_s,*})$ is achievable. Since
%
\begin{align*}
\frac{1}{l} I(\mvec{X}_{\epsilon,l}^{\phi_s};\mr{\mvec{X}}_{\epsilon,l}^{\phi_s,*}) = \ \inf_{\mmat{C}_{\mvec{S}_{\epsilon,l}^{\phi_s}}\in\mC_{\mvec{S}_{\epsilon,l}^{\phi_s}}}  \frac{1}{2l} \log\bigg(\frac{\det(\mmat{C}_{\mvec{X}_{\epsilon,l}^{\phi_s}})}{\det(\mmat{C}_{\mvec{S}_{\epsilon,l}^{\phi_s}})}\bigg),
      &
\end{align*}
    we conclude that 
\begin{multline}
    \!R_{\epsilon}^{\phi_s}(D) \!\leq 
    \limsup_{l\to\infty} \!\inf_{\mmat{C}_{\mvec{S}_{\epsilon,l}^{\phi_s}}\in\mC_{\mvec{S}_{\epsilon,l}^{\phi_s}}}\!   \frac{1}{2l} \log\bigg(\frac{\det(\mmat{C}_{\mvec{X}_{\epsilon,l}^{\phi_s}})}{\det(\mmat{C}_{\mvec{S}_{\epsilon,l}^{\phi_s}})}\bigg).\notag
\end{multline}
Therefore, Lemma~\ref{lem: achievability for given sampling phase} is proved.
\end{IEEEproof}


Combining Lemmas~\ref{lem: converse for given sampling phase} and \ref{lem: achievability for given sampling phase}, the \gls{rdf} for a given $\phi_s$ is obtained.
Lastly, for the scenario of Thm.~\ref{thm: main theorem}, the  equidistribution theorem (see \cite[Ex.~2.1]{kuipers1974} and \cite[Sec.~XI.1]{coppel2009}) implies that asymptotically, as the number of  generated source messages goes to infinity, the fact that the sampling interval $T_s(\epsilon)$ is incommensurate with the cyclostationarity period $T_c$ of the \gls{ct} \gls{wscs} source process, implies that the sampling phase $\phi_s$ of the generated source messages is uniformly distributed over the interval $[0,T_c)$. Accordingly, the \gls{rdf} $R_{\epsilon}(D)$ in Thm.~\ref{thm: main theorem} is obtained by averaging over $\phi_s$ as in Eqn.~\eqref{a 0140am}, which completes the proof of Thm.~\ref{thm: main theorem}.



\bibliographystyle{IEEEtran}
\bibliography{IEEEabrv, references.bib}

\end{document}